%% file: main.tex
\documentclass[11pt]{article}
\usepackage[margin=1in]{geometry}
\usepackage[utf8]{inputenc}
\usepackage[T1]{fontenc}
\usepackage{lmodern}
\usepackage{amsmath,amssymb}
\usepackage{booktabs}
\usepackage{threeparttable}
\usepackage{graphicx}
\usepackage{caption}
\usepackage{subcaption}
\usepackage{natbib}
\usepackage[hidelinks]{hyperref}
\usepackage{setspace}
\usepackage{enumitem}
\usepackage{float}
\usepackage{pdflscape}
\usepackage{adjustbox}
\input{cifras.tex}

\title{Discretion at the Margin:\\ Threshold Manipulation and Electoral Bans in Colombian Public Procurement\thanks{This version: October 2026. Comments welcome. All figures in this paper are generated from a verified claims registry; every number in the text is recomputed by an independent verification battery from the underlying data (see Appendix~\ref{app:datos}). Declarations of contributions, data availability and competing interests are given at the end of the paper.}}
\author{Cristian Espinal Maya\thanks{Corresponding author. Universidad Nacional de Colombia (PhD student in Engineering) and Universidad EAFIT. Email: cjespinalm@eafit.edu.co. ORCID: 0009-0000-1009-8388.}
\and Santiago Jim\'enez Londo\~no\thanks{School of Applied Sciences and Engineering, Universidad EAFIT. Email: sjimene8@eafit.edu.co.}}
\date{October 2026}

\begin{document}
\maketitle

\begin{abstract}
\noindent Public procurement laws trade rules against discretion with value thresholds: below them, buyers use faster and less competitive procedures. Colombia offers an unusually clean setting to measure how much buyers manipulate estimated values to stay below the line and what it costs. Thresholds are entity-specific---five budget bands set by law---and move every January with the minimum wage, while the round numbers that usually confound bunching estimates stay fixed. Using the universe of Colombian procurement records (SECOP I and II, 2011--2026; \DiagUnidades{} SECOP II processes) and each entity-year's own threshold, we find sharp bunching of estimated values just below both legal cutoffs: \Bminima{} excess processes below the small-purchase threshold (\Partminima{} of the mass observed in the last five percent below it) and \Bmenor{} below the tender threshold. The spike tracks the threshold, not the round numbers: it moves with the minimum wage, appears only at each entity's own band, and is an order of magnitude smaller for procedures that carry no value cap. A counterfactual built from entities of higher budget bands, at the same peso values and in the same year, gives the same order of magnitude. Territorial governments, run by elected mayors and governors, bunch more than appointed national agencies at the small-purchase threshold, even after reweighting for budget band, object and year. Processes bunched below the tender threshold attract half as many bidders as those just above it, obtain smaller discounts, and are more often enlarged after award; below the small-purchase threshold they go more often to recurrent suppliers. A second source of variation---the four-month ban on direct contracting before presidential elections---shows that buyers respond to rules on the timing margin: direct contracting changes by \LGDirectaVeda{} during the ban and rises \LGDirectaUltTwoEnero{} in the two weeks before it, in each of four electoral cycles, with a partial shift toward below-threshold procedures. Chile's fixed UTM thresholds display the same pattern, and a cutoff abolished in 2024 shows no bunching afterwards.
\end{abstract}


\newpage
\input{sec_intro.tex}
\input{sec_institucional.tex}
\input{sec_datos.tex}
\input{sec_estrategia.tex}
\input{sec_bunching.tex}
\input{sec_mecanismos.tex}

\input{sec_garantias.tex}
\input{sec_chile.tex}
\input{sec_conclusion.tex}

\input{sec_declaraciones.tex}

\bibliographystyle{aer}
\bibliography{referencias}

\clearpage
\appendix
\input{apendice.tex}
\end{document}

%% file: cifras.tex
\newcommand{\Bminima}{49,631}
\newcommand{\BEEminima}{2,076}
\newcommand{\bminima}{30.8}

\newcommand{\Partminima}{86\%}
\newcommand{\Nventminima}{167,349}
\newcommand{\Nexactominima}{12,561}
\newcommand{\Nentminima}{2,579}

\newcommand{\Bsinexactosminima}{37,070}
\newcommand{\BsinexactosEEminima}{1,524}
\newcommand{\BrobMinminima}{42,447}
\newcommand{\BrobMaxminima}{58,917}
\newcommand{\Bdibminima}{41,165}
\newcommand{\BdibEEminima}{1,614}

\newcommand{\Bmenor}{3,157}
\newcommand{\BEEmenor}{182}
\newcommand{\bmenor}{8.4}

\newcommand{\Partmenor}{63\%}
\newcommand{\Nventmenor}{43,337}
\newcommand{\Nexactomenor}{451}
\newcommand{\Nentmenor}{1,565}

\newcommand{\Bsinexactosmenor}{2,706}

\newcommand{\BrobMinmenor}{2,523}
\newcommand{\BrobMaxmenor}{3,932}
\newcommand{\Bdibmenor}{2,569}
\newcommand{\BdibEEmenor}{147}

\newcommand{\FracMultiploUnM}{28\%}

\newcommand{\PlSIRegimenEspecialminimab}{2.3}

\newcommand{\PlSIDirectaminimab}{1.4}
\newcommand{\SIBminima}{149,534}
\newcommand{\SIBEEminima}{6,702}
\newcommand{\SIPartminima}{83\%}

\newcommand{\SINminima}{488,908}

\newcommand{\PlSIRegimenEspecialmenorb}{0.4}

\newcommand{\PlSIDirectamenorb}{0.2}

\newcommand{\SubOrdenTerritorialminimab}{37.3}

\newcommand{\SubOrdenNacionalminimab}{22.3}

\newcommand{\SubOrdenCorporacionAutonomaminimab}{24.3}

\newcommand{\SubUmbralIndependienteCuipoPgnminimab}{19.0}
\newcommand{\SubUmbralIndependienteCuipoPgnminimabEE}{1.6}

\newcommand{\SubRedondoNoMultiplominimab}{31.6}

\newcommand{\SubVedaArtThreeThreeDentrominimab}{37.2}
\newcommand{\SubVedaArtThreeThreeDentrominimabEE}{1.7}

\newcommand{\SubVedaArtThreeThreeFueraminimab}{30.4}

\newcommand{\SubOrdenTramoTerritorialTwoEightZerominimab}{51.5}

\newcommand{\SubOrdenTramoTerritorialFourFiveZerominimab}{30.8}

\newcommand{\SubOrdenTramoNacionalTwoEightZerominimab}{27.2}

\newcommand{\SubOrdenTramoNacionalFourFiveZerominimab}{25.9}

\newcommand{\SubOrdenObjetoTerritorialBienesminimab}{38.9}

\newcommand{\SubOrdenObjetoTerritorialServiciosminimab}{35.8}

\newcommand{\SubOrdenObjetoNacionalBienesminimab}{22.6}

\newcommand{\SubOrdenObjetoNacionalServiciosminimab}{20.4}

\newcommand{\SubOrdenTerritorialmenorb}{10.2}

\newcommand{\SubOrdenNacionalmenorb}{6.3}

\newcommand{\SubObjetoObramenorb}{15.7}

\newcommand{\SubObjetoBienesmenorb}{4.6}

\newcommand{\SubObjetoConsultoriamenorB}{-15}
\newcommand{\SubObjetoConsultoriamenorBEE}{17}

\newcommand{\SubUmbralIndependienteCuipoPgnmenorb}{5.7}
\newcommand{\SubUmbralIndependienteCuipoPgnmenorbEE}{1.9}

\newcommand{\SubRedondoNoMultiplomenorb}{7.7}

\newcommand{\SubOrdenObjetoTerritorialObramenorb}{17.2}

\newcommand{\SubOrdenObjetoNacionalObramenorb}{12.1}

\newcommand{\FseB}{810}
\newcommand{\FseBEE}{85}
\newcommand{\Fseb}{11.2}
\newcommand{\FseN}{7,778}

\newcommand{\MasaminimaHOneFive}{12,327}
\newcommand{\MasaPctminimaHOneFive}{31\%}

\newcommand{\ResNProponentesminimaDif}{-0.23}

\newcommand{\ResPersonaNaturalminimaAm}{24.4\%}
\newcommand{\ResPersonaNaturalminimaArr}{12.8\%}

\newcommand{\ResRecurrenteminimaAm}{28.7\%}
\newcommand{\ResRecurrenteminimaArr}{19.3\%}
\newcommand{\ResRecurrenteminimaNoManip}{22.2\%}

\newcommand{\MasamenorHOneFive}{1,922}
\newcommand{\MasaPctmenorHOneFive}{80\%}
\newcommand{\ResNProponentesmenorAm}{3.52}
\newcommand{\ResNProponentesmenorArr}{6.97}

\newcommand{\ResNProponentesmenorDif}{-3.45}
\newcommand{\ResNProponentesmenorDifEE}{1.02}
\newcommand{\ResNProponentesmenorDifNM}{-0.20}
\newcommand{\ResNProponentesmenorDifNMEE}{0.53}

\newcommand{\ResUnSoloProponentemenorAm}{47.3\%}
\newcommand{\ResUnSoloProponentemenorArr}{39.1\%}

\newcommand{\ResDescuentomenorAm}{4.6\%}
\newcommand{\ResDescuentomenorArr}{8.7\%}

\newcommand{\ResDescuentomenorDif}{-4.2 pp}
\newcommand{\ResDescuentomenorDifEE}{1.1}
\newcommand{\ResDescuentomenorDifNM}{-0.8 pp}

\newcommand{\ResDescuentoContratomenorDif}{-4.4 pp}

\newcommand{\ResValorSobreUmbralmenorAm}{46.2\%}

\newcommand{\ResPAdicionValormenorAm}{48.9\%}
\newcommand{\ResPAdicionValormenorArr}{44.1\%}

\newcommand{\ResPAdicionValormenorDif}{+4.8 pp}
\newcommand{\ResPAdicionValormenorDifEE}{3.1}
\newcommand{\ResPAdicionValormenorDifNM}{+9.8 pp}
\newcommand{\ResPAdicionValormenorDifNMEE}{3.2}

\newcommand{\ResPAdicionTopemenorAm}{16.2\%}
\newcommand{\ResPAdicionTopemenorArr}{10.9\%}

\newcommand{\FraccTieneHermDif}{-0.4 pp}
\newcommand{\FraccTieneHermEE}{0.4}

\newcommand{\FraccHermSumaSobreUmbralDif}{+1.9 pp}

\newcommand{\FraccHermMismoProvDif}{+0.6 pp}
\newcommand{\FraccHermMismoProvEE}{0.1}
\newcommand{\FraccHermMismoProvLejos}{2.0\%}

\newcommand{\DiagFilas}{9,235,206}
\newcommand{\DiagIdsFase}{8,798,623}
\newcommand{\DiagUnidades}{8,569,218}
\newcommand{\DiagCompetitivos}{516,372}
\newcommand{\DiagCompetitivosConUmbral}{512,237}
\newcommand{\DiagEAconMinima}{15,016}
\newcommand{\DiagEAsospechosos}{97}
\newcommand{\DiagCobertura}{99\%}
\newcommand{\DiagIdsPorUnidadMenor}{2.9}
\newcommand{\DiagIdsPorUnidadLicitacion}{2.0}
\newcommand{\DiagIdsPorUnidadSubasta}{2.0}

\newcommand{\IdSmmlvAnteriorminimab}{-1.3}
\newcommand{\IdSmmlvSiguienteminimab}{-1.1}
\newcommand{\IdMatrizDiagminima}{32.4}
\newcommand{\IdMatrizFueraminima}{-0.9}
\newcommand{\FactorVentanaminima}{7.2}
\newcommand{\bRobMinminima}{14.4}
\newcommand{\bRobMaxminima}{54.3}
\newcommand{\BprimeraFasemenor}{3,161}
\newcommand{\IdSmmlvAnteriormenorb}{-0.6}
\newcommand{\IdSmmlvSiguientemenorb}{-0.9}
\newcommand{\IdMatrizDiagmenor}{7.6}
\newcommand{\IdMatrizFueramenor}{-0.4}
\newcommand{\FactorVentanamenor}{2.7}
\newcommand{\bRobMinmenor}{5.1}
\newcommand{\bRobMaxmenor}{11.0}
\newcommand{\RobSIModalminimab}{17.9}

\newcommand{\RobSIContemporaneominimab}{17.9}

\newcommand{\BEEtramoAniominima}{10,052}

\newcommand{\BtStatTramoAniominima}{4.9}
\newcommand{\RepondNacionalComoTerritorialminimab}{28.2}
\newcommand{\RepondNacionalComoTerritorialminimabEE}{2.5}
\newcommand{\RepondTerritorialObjetivominimab}{37.3}

\newcommand{\BEEtramoAniomenor}{624}

\newcommand{\BtStatTramoAniomenor}{5.1}
\newcommand{\RepondNacionalComoTerritorialmenorb}{8.3}
\newcommand{\RepondNacionalComoTerritorialmenorbEE}{1.1}
\newcommand{\RepondTerritorialObjetivomenorb}{10.2}

\newcommand{\RobSIN}{102,194}

\newcommand{\BpolyDosTramosminima}{34,223}
\newcommand{\SubOrdenTerritorialminimaPart}{88\%}
\newcommand{\SubOrdenTerritorialminimaBN}{35\%}
\newcommand{\SubOrdenNacionalminimaPart}{82\%}
\newcommand{\SubOrdenNacionalminimaBN}{21\%}

\newcommand{\BpolyDosTramosmenor}{2,093}

\newcommand{\NdibShare}{56\%}
\newcommand{\DineroExcesominimaBill}{2.9}

\newcommand{\MasaPctNacionalminimaHOneFive}{44\%}

\newcommand{\MasaPctTerritorialminimaHOneFive}{31\%}

\newcommand{\DineroExcesomenorBill}{2.0}

\newcommand{\MasaPctNacionalmenorHOneFive}{74\%}

\newcommand{\MasaPctTerritorialmenorHOneFive}{99\%}

\newcommand{\DineroDescuentoPerdidoMilM}{84}

\newcommand{\LGDirectaVeda}{-95\%}
\newcommand{\LGDirectaVedaP}{0.001}
\newcommand{\LGDirectaAnticipacion}{+29\%}
\newcommand{\LGDirectaAnticipacionP}{0.181}
\newcommand{\LGDirectaUltTwoEnero}{+209\%}
\newcommand{\LGDirectaUltTwoEneroP}{0.001}
\newcommand{\LGDirectaRebote}{+6\%}
\newcommand{\LGDirectaReboteP}{0.872}
\newcommand{\LGDirectaNeto}{-26\%}
\newcommand{\LGDirectaNetoP}{0.177}
\newcommand{\LGMinimaVeda}{+18\%}
\newcommand{\LGMinimaVedaP}{0.005}

\newcommand{\LGMenorVeda}{+21\%}

\newcommand{\LGLicitacionVeda}{+56\%}

\newcommand{\LGRegimenEspecialVeda}{-46\%}

\newcommand{\LGdirectaVedaTwoZeroOneFour}{-2.95}
\newcommand{\LGdirectaVedaTwoZeroOneEight}{-2.61}
\newcommand{\LGdirectaVedaTwoZeroTwoTwo}{-2.94}
\newcommand{\LGdirectaVedaTwoZeroTwoSix}{-3.85}
\newcommand{\LGIIDirectaExceptuadasVeda}{-20\%}
\newcommand{\LGIIDirectaExceptuadasVedaP}{0.679}
\newcommand{\LGIIDirectaNoExceptuadasVeda}{-94\%}
\newcommand{\LGIIDirectaNoExceptuadasVedaP}{0.036}

\newcommand{\PtresDD}{0.34}
\newcommand{\PtresP}{0.43}
\newcommand{\PtresDDctrl}{0.40}
\newcommand{\LGartThreeEightConveniosTerritorialesVeda}{-88\%}
\newcommand{\LGartThreeEightConveniosTerritorialesVedaP}{0.05}
\newcommand{\LGartThreeEightConveniosNacionalesVeda}{-90\%}
\newcommand{\LGartThreeEightConveniosNacionalesVedaP}{0.14}
\newcommand{\ChlUOneZeroZeroVisibleB}{27,150}
\newcommand{\ChlUOneZeroZeroVisibleBEE}{3,822}
\newcommand{\ChlUOneZeroZeroVisibleb}{11.7}

\newcommand{\ChlUOneZeroZeroZeroVisibleB}{5,969}

\newcommand{\ChlUOneZeroZeroZeroVisibleb}{7.4}

\newcommand{\ChlUTwoZeroZeroZeroPlaceboTwoZeroTwoFiveTwoZeroTwoSixB}{-49}
\newcommand{\ChlUTwoZeroZeroZeroPlaceboTwoZeroTwoFiveTwoZeroTwoSixBEE}{31}

\newcommand{\ChlMuestraN}{1,642,250}
\newcommand{\ChlOrganismos}{1,179}
\newcommand{\ChlCApre}{4,775}
\newcommand{\ChlCApost}{837}
\newcommand{\ChlCAcaida}{82\%}
\newcommand{\PolMargenminimaTOneb}{43.9}
\newcommand{\PolMargenminimaTOnebEE}{2.7}
\newcommand{\PolMargenminimaTTwob}{50.0}

\newcommand{\PolMargenminimaTThreeb}{51.2}

\newcommand{\PolMargenminimaCerradaFiveppb}{44.6}

\newcommand{\PolMargenminimaNoCerradab}{50.5}

\newcommand{\PolMargenmenorTOneb}{12.8}

\newcommand{\PolMargenmenorTTwob}{14.4}

\newcommand{\PolMargenmenorTThreeb}{10.8}

\newcommand{\PolMargenmenorCerradaFiveppb}{13.0}

\newcommand{\PolMargenmenorNoCerradab}{11.6}

\newcommand{\PolDonmenorAmontonado}{0.76\%}

\newcommand{\PolDonmenorLejos}{0.47\%}

\newcommand{\PolProcesosAlcaldias}{1,167,833}
\newcommand{\PolCruzados}{887,371}
\newcommand{\PolMunicipios}{589}

%% file: sec_intro.tex
\section{Introduction}\label{sec:intro}

Governments buy about a third of what they spend, and the rules that govern how they buy are written around value thresholds. Below a cutoff, a buyer may use a faster procedure with fewer bidders and less publicity; above it, an open tender. The design trades the cost of rules against the risk of discretion \citep{Bandiera2009, Coviello2018, Decarolis2025}, and a growing literature shows that buyers respond to the cutoff by manipulating the estimated value of what they buy \citep{PalgutaPertold2017, Szucs2024, Coviello2026}. How much they do so, who does it, and what it costs are open questions with a practical edge: every procurement reform that moves a threshold assumes an answer to them.

This paper measures threshold manipulation in Colombia using the complete universe of public procurement records for 2011--2026 and a source of variation that other settings lack. Colombian thresholds are set by law as a function of each entity's annual budget, in five bands expressed in legal monthly minimum wages: the tender threshold ranges from 280 to 1,000 minimum wages depending on the band, and the small-purchase threshold is ten percent of it. Because the minimum wage is reset every January, the cutoff in pesos moves each year---by between 3.5 and 23 percent---for every entity, while the round numbers at which buyers naturally set budgets stay put. A spike in the density of estimated values that is caused by the threshold must follow the threshold across bands and years; a spike caused by rounding must not. We use the cross-sectional variation to build a counterfactual that does not rely on a global polynomial---for entities of a given band, the density of estimated values of entities in other bands, at the same peso values and in the same year, a design that \citet{IvarsOlave2025} apply to national entities and that \citet{Kleven2016} describes as difference-in-bunching---and the temporal variation to construct placebos that should, and do, give zero.

We find sharp and large bunching. Just below the small-purchase threshold there are \Bminima{} more processes than a smooth density would place there, \Partminima{} of all the processes observed in the last five percent below the cutoff, with an estimated value of about COP \DineroExcesominimaBill{} trillion (US\$0.7 billion) over 2015--2026; below the tender threshold, \Bmenor{} processes (\Partmenor), worth COP \DineroExcesomenorBill{} trillion. The excess vanishes when the threshold is placed where it was the previous year, appears only at each entity's own band, and is absent for procedures with no value cap---consulting contracts, which go to a merit contest at any value, show none. The cross-band counterfactual gives the same numbers within 1.5 standard errors. The same pattern appears in the earlier platform (SECOP~I, 2011--2020) and at a third cutoff, the 20-minimum-wage limit below which public schools contract under their own rules.

Who manipulates is informative about why. Territorial governments---run by elected mayors and governors---bunch more than appointed national agencies at the small-purchase threshold, and the gap survives reweighting national agencies to the territorial mix of budget bands, objects and years; at the tender threshold the two are statistically indistinguishable. This is the reverse of the pattern documented for Italy by \citet{Coviello2026}, where appointed administrations manipulate and elected ones do not. Two features of the Colombian case can explain the reversal: mayors and governors cannot be immediately re-elected, so the electoral discipline that Coviello et al.\ invoke is absent by construction, and Colombian campaign donors are repaid with small-purchase contracts \citep{Ruiz2018, Gulzar2025}. Consistent with the first, the mayor's margin of victory does not predict bunching. Bunched processes are also different from their neighbors. Recovering the outcomes of the excess mass with the decomposition of \citet{DiamondPersson2016}, we find that processes bunched below the tender threshold attract \ResNProponentesmenorAm{} bidders against \ResNProponentesmenorArr{} for processes just above it, obtain a discount of \ResDescuentomenorAm{} against \ResDescuentomenorArr---a gap worth about COP \DineroDescuentoPerdidoMilM{} billion over the period---and receive post-award additions more often; \ResValorSobreUmbralmenorAm{} of them end above the threshold they avoided. Below the small-purchase threshold the number of bidders is barely affected, and the difference is in who wins: bunched processes go to recurrent suppliers of the entity and to individuals rather than firms.

A second rule lets us see the timing margin of discretion. Law 996 of 2005 bans direct contracting by all public entities during the four months before a presidential election. Pooling both platforms across four electoral cycles, direct contracting changes by \LGDirectaVeda{} during the ban---compliance is close to complete---and, more informatively, rises \LGDirectaUltTwoEnero{} in the two weeks before it, in every cycle, with no rebound afterwards and a net change over the cycle that we cannot distinguish from zero. Small purchases rise \LGMinimaVeda{} during the ban: part of the activity the rule closes moves toward the procedure whose threshold buyers manipulate.

Finally, Chile provides a comparison with thresholds that do not depend on the buyer and are denominated in an indexed tax unit. With the same estimator and the same sample rules, Chilean tenders bunch below the 100 and 1,000 UTM cutoffs ($b = \ChlUOneZeroZeroVisibleb$ and $\ChlUOneZeroZeroZeroVisibleb$), and a cutoff that the 2024 reform abolished shows no bunching afterwards. The excess relative to the counterfactual bin is larger in Colombia at the lower cutoff and similar at the upper one.

\paragraph{Contribution.} We contribute to three literatures. First, to the measurement of threshold manipulation in procurement \citep{PalgutaPertold2017, Szucs2024, Coviello2026, Carril2026, Fazio2022, Castro2026}. Our setting supplies within-country, within-entity variation in the cutoff, which lets us separate the threshold from focal points and build a counterfactual from other entities; our sample covers national and territorial buyers on both platforms over fifteen years. The closest paper is \citet{IvarsOlave2025}, who use SECOP~II data for national entities in 2018--2024 with the neighboring budget band as control and find that at the small-purchase threshold the excess below is matched by the missing mass above (``shifting''). We cover both levels of government and both platforms, add the temporal variation of the minimum wage, placebos with a strict zero, the outcomes of bunched processes, and the electoral ban; and we reconcile the two findings: for national entities the missing mass is \MasaPctNacionalminimaHOneFive{} of the excess, as they find, whereas for territorial entities it is \MasaPctTerritorialminimaHOneFive, so the extensive margin is a territorial phenomenon. Second, to the literature on discretion and political accountability in procurement \citep{Bandiera2009, Best2023, Decarolis2025, Gerardino2024}: the elected-versus-appointed contrast with Italy, and the interaction between the electoral ban and the thresholds, speak to when discretion is disciplined and when it is not. Third, to the empirical bunching literature \citep{KlevenWaseem2013, Kleven2016}: the cross-band design is an application of the difference-in-bunching idea to a setting where the notch varies across agents and over time.

The rest of the paper proceeds as follows. Section~\ref{sec:institucional} describes the thresholds and the electoral ban. Section~\ref{sec:datos} describes the data and the unit of analysis. Section~\ref{sec:estrategia} sets out the estimators. Section~\ref{sec:bunching} presents the bunching results and their validation. Section~\ref{sec:mecanismos} examines where the excess comes from and what happens to bunched processes. Section~\ref{sec:garantias} studies the electoral ban. Section~\ref{sec:chile} compares with Chile. Section~\ref{sec:conclusion} discusses limitations and policy implications.

%% file: sec_institucional.tex
\section{Institutional background}\label{sec:institucional}

\subsection{Procedures and thresholds}

Colombian public procurement under the general statute (Law 80 of 1993 as amended by Law 1150 of 2007) offers five competitive procedures and direct contracting. The open tender (\emph{licitaci\'on p\'ublica}) is the default. Three procedures are defined by the nature of the object, not by value: the reverse auction for goods and services of uniform technical characteristics, the merit contest for consulting, and direct contracting for a closed list of causes (professional services, sole suppliers, intergovernmental agreements, emergencies). Two are defined by value. The abbreviated selection of lesser value (\emph{selecci\'on abreviada de menor cuant\'ia}) replaces the tender for contracts up to the entity's ``menor cuant\'ia'' threshold; it requires public notice, an expression-of-interest stage and an evaluation, but no public hearing and shorter terms. The small-purchase procedure (\emph{m\'inima cuant\'ia}, Law 1474 of 2011, art.~94) applies, ``regardless of the object,'' to contracts whose value does not exceed ten percent of that threshold: a five-day invitation, award to the lowest price, no guarantees required. Below the small-purchase threshold every object goes through this procedure; above it, the object determines whether the buyer runs an abbreviated selection, a reverse auction or a merit contest.

The tender threshold is set in Law 1150 (art.~2, num.~2, lit.~b) as a step function of the entity's annual budget expressed in legal monthly minimum wages (SMMLV): 1,000 SMMLV for budgets of 1.2 million SMMLV or more; 850 for budgets between 850,000 and 1.2 million; 650 between 400,000 and 850,000; 450 between 120,000 and 400,000; and 280 below 120,000. The provision has never been amended. The small-purchase threshold is ten percent of it. Because the minimum wage is set by decree every December for the following year, both thresholds move in pesos every January; between 2010 and 2026 the annual adjustment ranged from 3.5 to 23 percent. In 2024, for instance, an entity in the lowest band could award by abbreviated selection up to COP 364 million and by small purchase up to COP 36.4 million; in 2025 the limits were 398.6 and 39.9 million. Public schools contract through a separate regime: below 20 SMMLV they follow the rules of their own governing council (Decree 1075 of 2015, art.~2.3.1.6.3.17), and above it the general statute.

Two further rules matter for interpretation. Contracts may be increased after award by up to fifty percent of their initial value (Law 80, art.~40, paragraph), so a buyer who stays below a threshold at award is not bound by it afterwards. And a large set of entities---public universities, state-owned enterprises, utilities, some financial and health institutions---operate under special regimes with their own manuals; their processes appear in the data under a ``special regime'' procedure and, when they use general-statute procedures, are bound by the same thresholds.

\subsection{The electoral ban}

Law 996 of 2005, the ``Guarantees Law,'' restricts the use of public resources around elections. Its article~33 prohibits direct contracting by ``all entities of the State'' during the four months before a presidential election and until the second round, exempting defense and security, public credit, emergencies, the reconstruction of infrastructure after attacks or disasters, and health and hospital entities. Every other procedure remains available. The ban applied before the presidential elections of 2010, 2014, 2018, 2022 and 2026; in 2026 it ran from 31 January to 21 June. A milder restriction (article~38, paragraph) prevents territorial governments from signing intergovernmental agreements in the four months before any election, including local ones; a 2021 budget law lifted it for 2022 and the Constitutional Court struck the provision down with retroactive effect on 5 May 2022 (ruling C-153 of 2022). The Council of State has held that both restrictions bind territorial entities.

\subsection{The procurement platforms}

Every process is published on the Electronic System for Public Procurement (SECOP). SECOP~I, in use since 2011, is a publication platform: entities upload documents and register the contract. SECOP~II, launched in 2015, is transactional: the process, the offers, the award, the contract and its modifications are created on the platform. Its use became mandatory for national entities and departmental capitals in 2020, for special-regime entities in July 2022, for municipalities in staggered cut-offs during 2022, and for the remaining entities in 2025. For municipal buyers before 2022, therefore, SECOP~I is the exhaustive source and SECOP~II a self-selected subset; we use SECOP~I for the 2011--2020 replication and SECOP~II, whose estimated values and bidder lists are cleaner, for the main analysis.

\subsection{Chile}

Chile's Public Procurement Law (Law 19.886 and its regulation) classifies tenders by their estimated value in monthly tax units (UTM, about COP 290,000 or US\$70): L1 below 100 UTM, LE between 100 and 1,000, LP above 1,000, with two further bands (LQ, 2,000--5,000; LR, 5,000 and above) introduced in 2015 and partly removed by the 2024 regulation, which merged LQ into LP. Each band imposes longer publication terms and additional requirements. Direct purchases through ``Compra \'Agil'' were allowed below 10 UTM, then 30 UTM from January 2020, and 100 UTM from 12 December 2024 under Law 21.634. Unlike Colombia's, the thresholds do not depend on the buyer, and they move only with the monthly indexation of the UTM.

%% file: sec_datos.tex
\section{Data}\label{sec:datos}

\subsection{Procurement records}

We use the complete open-data universe of Colombian public procurement published by the national procurement agency (Colombia Compra Eficiente) on the national open-data portal. Two platforms coexist. SECOP~I (2011--2019, with residual use afterwards) records each contract once, with the estimated value of the process (\emph{cuant\'ia del proceso}), the procedure, the buyer and the contractor. SECOP~II (2015--2026, mandatory for national entities since 2020 and progressively for territorial ones) is transactional: it records the process with its estimated value (\emph{precio base}), the list of bidders, the bids, the award, the contract, its modifications and its payments. Because SECOP~II became mandatory for municipalities only in 2022 (Section~\ref{sec:institucional}), its territorial coverage before that year is partial and self-selected; the SECOP~I replication covers those years exhaustively. We downloaded every table page by page, verified each page against the server's own row counts, and archived the raw files with cryptographic hashes; Appendix~\ref{app:datos} documents the pipeline and the reconciliation of each table.

Two features of SECOP~II matter for the unit of analysis. First, a multi-stage procedure (abbreviated selection, tender, reverse auction, merit contest) appears as two or three records, one per stage (comments on the draft, expression of interest, submission of offers), each with its own process identifier. What ties them together is the portfolio identifier, which is also the key that links processes to contracts and bids. We therefore define the unit as the portfolio, keep the record of the most advanced stage (its estimated value and identifiers), and date the process at its first publication; using instead the estimated value of the earliest stage, which is the one that fixes the procedure, changes the excess mass by less than one percent (\BprimeraFasemenor{} against \Bmenor{} at the tender threshold). Ignoring this would count each abbreviated selection 2.9 times and each tender twice, and would mechanically overstate the density just above the small-purchase threshold. Second, the process-level award flag is unreliable outside small purchases; we measure awards and outcomes from the contract table instead.

The SECOP~II process table contains \DiagFilas{} rows, \DiagIdsFase{} distinct stage identifiers (the difference is repeated rows from paginating a live dataset, which we report and drop) and \DiagUnidades{} processes; an abbreviated selection has \DiagIdsPorUnidadMenor{} stage identifiers on average, a tender \DiagIdsPorUnidadLicitacion, a reverse auction \DiagIdsPorUnidadSubasta. Of these processes, \DiagCompetitivos{} were run through one of the five competitive procedures of the general statute between 2015 and 2026 with an estimated value above zero, and we match \DiagCompetitivosConUmbral{} of them (\DiagCobertura) to their entity-year threshold.

\subsection{Thresholds}\label{sec:umbrales}

The tender threshold (\emph{menor cuant\'ia}) of an entity is set by law as a function of its annual budget expressed in legal monthly minimum wages (SMMLV): 1,000 SMMLV for entities with budgets of 1.2 million SMMLV or more, then 850, 650, 450 and 280 SMMLV for successively smaller budgets (Section~\ref{sec:institucional}). The small-purchase threshold (\emph{m\'inima cuant\'ia}) is ten percent of it. Because the minimum wage is reset every January, the peso value of every threshold moves each year---by between 3.5 and 23 percent---while the band structure is fixed.

We recover each entity-year's band from three sources, in order of priority. (i) The annual procurement plan (PAA) that every entity publishes in SECOP~II declares its tender and small-purchase thresholds; we accept a declared threshold only if it lies within three percent of a legal band for that year (a rule that also catches the two fields being swapped). (ii) For territorial entities from 2023 on, the initial appropriation of the expenditure budget reported to the national accounting system (CUIPO), linked to the entity by its tax identifier. (iii) For national entities from 2019 on, the initial appropriation in the national budget (PGN), matched by name. When an entity-year has none of these, we impute the modal band of the same entity in other years and flag it. Where a self-declared and a budget-based band coexist, they agree in 92 percent of entity-years. Self-declared thresholds account for 93 percent of the estimation sample; Section~\ref{sec:robustez} shows that the results hold on the budget-based subsample alone. Finally, we drop the \DiagEAsospechosos{} entity-years (of \DiagEAconMinima) in which more than five percent of the processes labelled as small purchases exceed the recovered small-purchase threshold, a sign that the band is wrong.

\subsection{Other sources}

We add: the SECOP~I universe (12.4 million records, 2011--2020), for which the threshold is the entity's modal band; the list of bidders per process and the bids (47.7 million); contract modifications; the mercantile registry (RUES) for supplier age and registration; the 2019 campaign-finance records of the National Electoral Council (Cuentas Claras) for donors to mayoral campaigns; and mesa-level electoral results from the Registrar's office for the 2019 and 2023 local elections. For Chile we use the complete Mercado P\'ublico archive of tenders (2015--2026), one record per tender, and the monthly value of the tax unit (UTM) in which Chilean thresholds are denominated.

\subsection{Sample for the bunching analysis}

The estimation sample is threshold-invariant by construction: it includes every process run under any of the competitive procedures of the general statute---small purchases, abbreviated selection of lesser value, reverse auctions, merit contests and open tenders---whatever its value. This matters because below the small-purchase threshold every object goes through the small-purchase procedure, whereas above it uniform goods go to reverse auctions and consulting to merit contests; a sample restricted to the value-capped procedures changes composition exactly at the cutoff and overstates the missing mass above it. Direct contracting and special-regime procedures are excluded from the estimation sample because their choice does not depend on value; they serve as placebos. Table~\ref{tab:datos} summarizes the samples.

\begin{table}[H]\centering\small
\caption{Samples}\label{tab:datos}
\begin{tabular}{lrr}
\toprule
 & Processes & Entities \\
\midrule
SECOP II stage identifiers, 2015--2026 & \DiagIdsFase{} & \\
SECOP II processes (portfolios) & \DiagUnidades{} & \\
Competitive procedures (five) with estimated value $>0$ & \DiagCompetitivos{} & \\
\quad with a recovered entity-year threshold & \DiagCompetitivosConUmbral{} & \\
Within $(0.5, 1.5]$ of the small-purchase threshold & \Nventminima{} & \Nentminima{} \\
Within $(0.5, 1.5]$ of the tender threshold & \Nventmenor{} & \Nentmenor{} \\
SECOP I, 2011--2020, within $(0.5, 1.5]$ of the small-purchase threshold & \SINminima{} & \\
Chile, 2015--2026, tenders in Chilean pesos with a positive estimated amount & \ChlMuestraN{} & \ChlOrganismos{} \\
\bottomrule
\end{tabular}
\end{table}

%% file: sec_estrategia.tex
\section{Empirical strategy}\label{sec:estrategia}

\subsection{Bunching at entity-year thresholds}

Let $v_i$ be the estimated value of process $i$ and $u_{e(i),t(i)}$ the tender threshold of its entity in its year. We work with the normalized value $r_i = v_i / u_{e,t}$, so that the tender cutoff is at $r=1$ and the small-purchase cutoff at $r=0.1$; for each cutoff $c \in \{0.1, 1\}$ we define $z_i = r_i/c$ and study the density of $z$ on $(0.5, 1.5]$ in bins of width one percent of the threshold, with right-closed intervals so that a value exactly equal to the threshold---which the law allows---falls in the last bin below it.

Following \citet{KlevenWaseem2013} and \citet{Kleven2016}, the counterfactual density is a polynomial fitted to the bin counts outside an excluded window $(z_L, z_U]$ around the cutoff:
\begin{equation}
n_j = \sum_{k=0}^{p} \beta_k (z_j - 1)^k + \sum_{l \in (z_L, z_U]} \gamma_l \mathbf{1}[j = l] + \varepsilon_j, \qquad \hat c_j = \sum_k \hat\beta_k (z_j-1)^k .
\end{equation}
The excess mass below the cutoff is $B = \sum_{z_L < z_j \le 1} (n_j - \hat c_j)$ and the missing mass above it is $M = \sum_{1 < z_j \le z_U} (\hat c_j - n_j)$. We report $B$ in number of processes, its share of the observed mass in $(z_L, 1]$, and---for comparability with the literature---$b = B / \overline{\hat c}_{(z_L,1]}$, the excess relative to the average counterfactual bin. Our preferred specification fixes $z_L = 0.95$ and $z_U = 1.04$ and chooses the polynomial degree (between 2 and 10) by leave-one-out cross-validation on the bins outside a slightly wider window, $(0.95, 1.10]$; the chosen degree is 9 at the small-purchase cutoff and 10 at the tender cutoff, and Appendix~\ref{app:robustez} shows that $B$ moves by less than a fifth across degrees 3 to 10 and across windows. Standard errors come from a block bootstrap that resamples entities with replacement, holding the degree fixed; they capture the correlation of values within a buyer but not shocks common to a band-year. We also report the wider window at which the missing mass above would balance the excess below, capped at $z_U = 1.25$; at the small-purchase cutoff the cap binds, which is itself informative (Section~\ref{sec:mecanismos}).

Three features of the setting make this design more credible than in a single-threshold country. First, the cutoff in pesos differs across the five budget bands and moves every January, while round-number focal points (multiples of one million pesos, \FracMultiploUnM{} of the sample) do not: a spike caused by the threshold must follow it, and a spike caused by rounding must not. Second, the same estimator applied to procedures that carry no value cap---reverse auctions and merit contests at the tender cutoff, direct contracting and special-regime procedures at both---should find nothing. Third, the band structure allows a counterfactual that does not rely on a global polynomial.

\subsection{Cross-band counterfactual}\label{sec:dib}

For entities in band $T$ in year $t$, the threshold in pesos is $u_{T,t} = T \times \mathrm{SMMLV}_t$. Entities in a higher band $T' > T$---so that their own cutoff, at $u_{T',t}/u_{T,t} \ge 1.89$, and the missing mass above it lie entirely outside the window $(0.5, 1.5]$---face no incentive to bunch at $u_{T,t}$. Lower bands cannot serve as controls: the hole above their own cutoff, which extends to about 1.5 times it, would fall inside the region where the counterfactual is fitted. Their density of estimated values, measured in the same pesos and the same year, is a counterfactual for band $T$ that embeds the round-number spikes, the calendar and the inflation of that year. Because entities of different bands differ in size, we rescale the control density with a quadratic polynomial fitted to the log ratio of treated to control counts outside the window $(0.90, 1.10]$, and compute $B$ and $M$ as before. Standard errors resample treated and control entities separately. The design is available for the two lowest bands, 280 (controls 650, 850 and 1000) and 450 (controls 850 and 1000), which contain \NdibShare{} of the processes in the estimation window. When we trace the missing mass further above the cutoff (Section~\ref{sec:mecanismos}) we use the same controls, a linear fit of the log ratio outside $(0.90, 1.50]$, and the explicit assumption that no mass is missing beyond 1.5 times the cutoff.

\subsection{Outcomes of bunched processes}\label{sec:dp}

Comparing outcomes just below and just above the threshold does not identify the effect of the procedure, because the processes just below are a mixture of non-manipulated processes and manipulated ones that would otherwise be above. Following \citet{DiamondPersson2016}, we recover the average outcome of the excess mass. Let $\bar y_j$ be the mean outcome in bin $j$ and $\hat y_j$ a quadratic fit of the bin means on $z$ over $(0.6, z_L]$, extrapolated into the window. The mean outcome among bunched processes is
\begin{equation}
\bar y_B = \frac{\sum_{z_L < z_j \le 1} n_j \bar y_j - \sum_{z_L < z_j \le 1} \hat c_j \hat y_j}{B},
\end{equation}
which we compare with $\hat y$ (non-manipulated processes of the same value, under the same procedure) and with the outcome extrapolated from above the cutoff (what the same processes would face under the more demanding procedure). The comparison is descriptive: the bunchers are selected. Standard errors resample entities.

\subsection{Electoral bans}\label{sec:est_lg}

Article~33 of Law 996 of 2005 bans direct contracting by every public entity during the four months before a presidential election and until the second round. We build a weekly series of signed contracts by procedure, pooling SECOP~I and SECOP~II (the platforms do not overlap for a given contract, and pooling avoids the migration between them), for 2012--2026 excluding 2020. For each year we align weeks on the start of the ban (in presidential years) or on the same calendar date (in the others), normalize by the mean of weeks $-20$ to $-9$, and take the difference in log counts between the four presidential years (2014, 2018, 2022, 2026) and the ten non-election years. We summarize the difference over the last two weeks before the ban, the eight weeks before it, the ban itself (weeks 0--16), the weeks after the second round (21--30) and the net change from week $-8$ to $+30$. With four treated years, asymptotic inference is not credible; $p$-values come from exact randomization inference over all $\binom{14}{4}=1{,}001$ assignments of four ``presidential'' years. SECOP~II alone allows a triple difference with the sectors the law exempts (health and defense).

%% file: sec_bunching.tex
\section{Bunching below the thresholds}\label{sec:bunching}

\subsection{Main estimates}

\begin{figure}[H]\centering
\includegraphics[width=\textwidth]{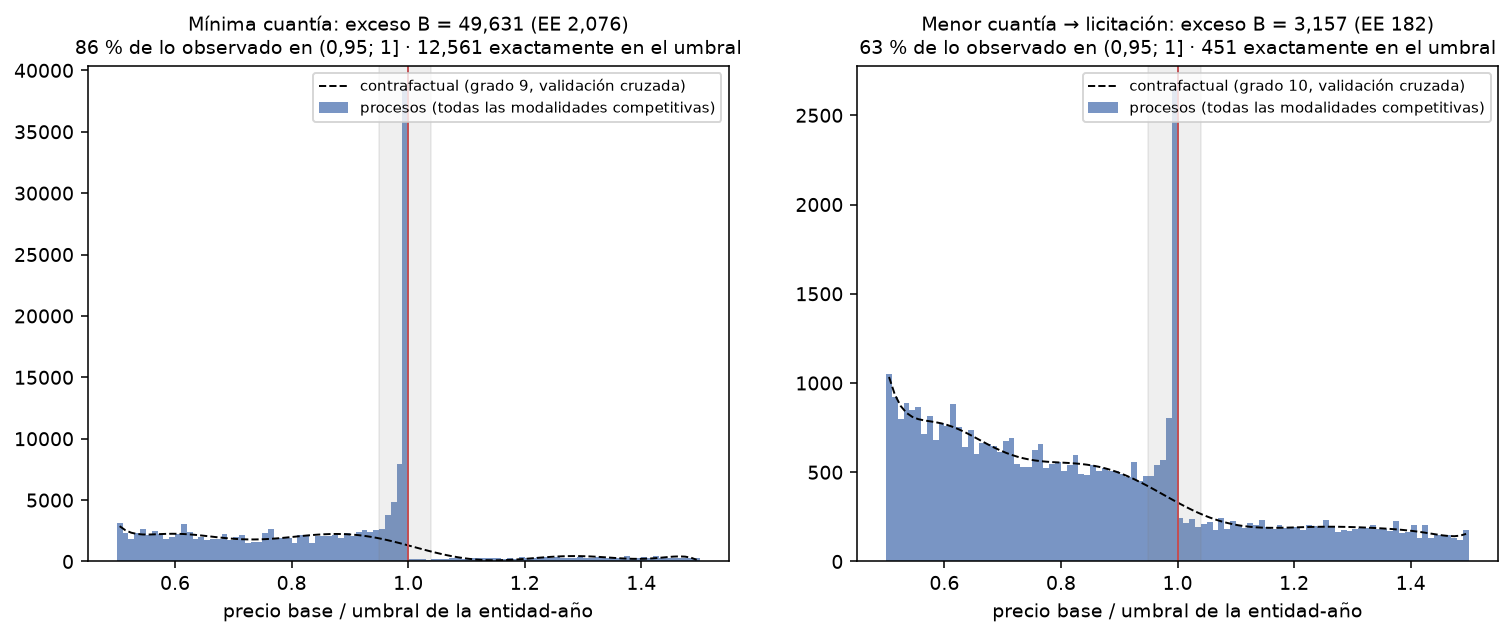}
\caption{Density of estimated values relative to the entity-year threshold. All competitive procedures of the general statute, SECOP~II 2015--2026; bins of one percent of the threshold; dashed line: polynomial counterfactual fitted outside the shaded window; degree chosen by cross-validation.}\label{fig:principal}
\end{figure}

Figure~\ref{fig:principal} plots the density of estimated values relative to each entity-year's own threshold, for all competitive procedures. At both cutoffs the density is flat or smoothly declining until about five percent below the threshold, rises sharply in the last bins, and collapses just above it. The excess mass is \Bminima{} processes (s.e.\ \BEEminima) below the small-purchase threshold and \Bmenor{} (s.e.\ \BEEmenor) below the tender threshold; it represents \Partminima{} and \Partmenor{} of all the processes observed in the last five percent below each cutoff (Table~\ref{tab:principal}). In the units of the literature, $b = \bminima$ and $\bmenor$: the excess equals 31 and 8 times the counterfactual mass of a single bin, or, cumulatively, the last five bins contain \FactorVentanaminima{} and \FactorVentanamenor{} times the mass that a smooth density would place there. Of the excess at the small-purchase threshold, \Nexactominima{} processes have an estimated value \emph{exactly} equal to the threshold, which the law allows; excluding them leaves $B = \Bsinexactosminima$ (s.e.\ \BsinexactosEEminima). At the tender threshold \Nexactomenor{} are exact and $B$ without them is \Bsinexactosmenor.

\begin{table}[H]\centering\small
\caption{Excess mass below the thresholds}\label{tab:principal}
\begin{adjustbox}{max width=\textwidth}
\input{tablas/principal.tex}
\end{adjustbox}
\begin{minipage}{\textwidth}\footnotesize Standard errors from a block bootstrap over entities (300 replications). ``Share of bin mass'' is $B$ divided by the observed number of processes in $(0.95, 1]$. The integration-constraint row widens $z_U$ until the missing mass above equals the excess below, up to 1.25. The cross-band row uses the counterfactual of Section~\ref{sec:dib}.\end{minipage}
\end{table}

The excess is stable across specifications: over a grid of five polynomial degrees, four lower limits and four upper limits of the window (80 specifications, Appendix Table~\ref{tab:robustez}), $B$ ranges from \BrobMinminima{} to \BrobMaxminima{} at the small-purchase threshold and from \BrobMinmenor{} to \BrobMaxmenor{} at the tender threshold. Standard errors that resample band-year blocks instead of entities, and so allow for shocks common to a band in a year (for instance, budgeting habits tied to that year's minimum wage), are about five times larger---\BEEtramoAniominima{} for $B$ at the small-purchase threshold and \BEEtramoAniomenor{} at the tender threshold---but leave both estimates \BtStatTramoAniominima{} and \BtStatTramoAniomenor{} standard errors from zero. The ratio $b$ is more sensitive (\bRobMinminima{} to \bRobMaxminima{} and \bRobMinmenor{} to \bRobMaxmenor{} over the same grid), because it divides by a counterfactual bin that the polynomial pins down less precisely and that shrinks with the degree; cross-validation favors high degrees, which is why we lead with $B$ and the share of bin mass and report $b$ only for comparability.

\subsection{The spike follows the threshold}

The identifying variation is that the threshold moves and the round numbers do not. Three exercises use it.

\paragraph{Wrong-year thresholds.} Recomputing $r$ with the previous year's minimum wage---so that the cutoff is placed where it was twelve months earlier---yields $b = \IdSmmlvAnteriorminimab$ at the small-purchase cutoff and $\IdSmmlvAnteriormenorb$ at the tender cutoff; with the next year's wage, $\IdSmmlvSiguienteminimab$ and $\IdSmmlvSiguientemenorb$. This is a test of location rather than of existence: displacing the cutoff by one year's wage adjustment (3.5 to 23 percent) moves the true spike out of the estimation window, and the estimator finds nothing at the displaced point. Figure~\ref{fig:movimiento} shows the same fact in pesos: within the 280 SMMLV band, the spike sits at each year's threshold and moves with it, while the round-number spikes stay put.

\begin{figure}[H]\centering
\includegraphics[width=0.85\textwidth]{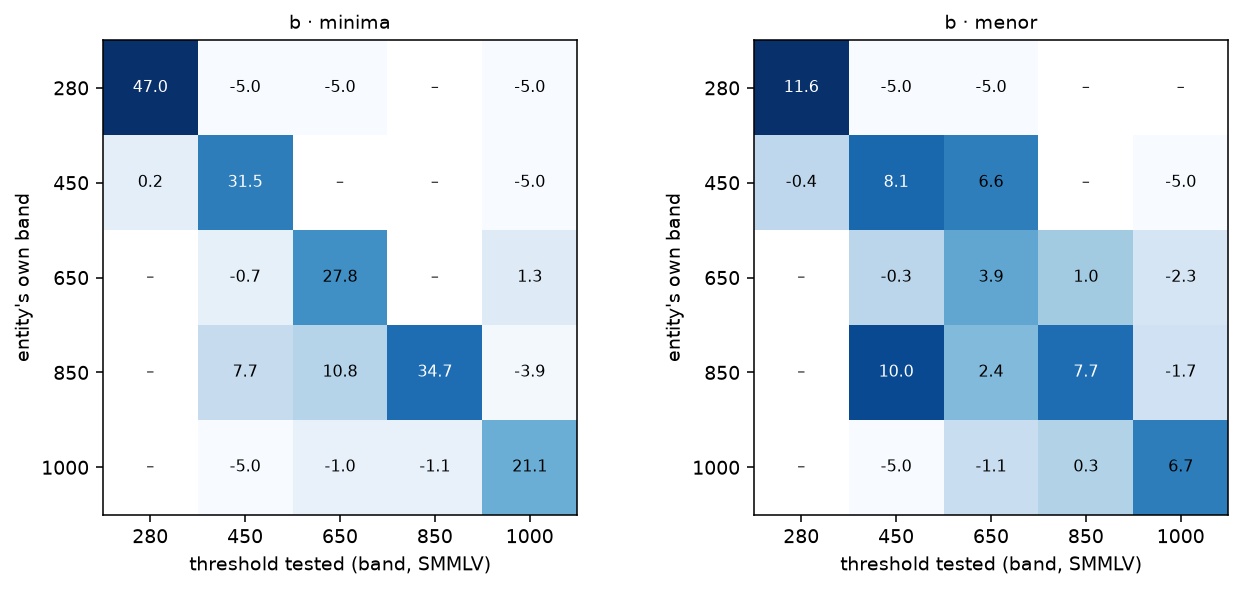}
\caption{Bunching at own and other bands' thresholds. Rows: true budget band of the entity; columns: threshold at which bunching is estimated.}\label{fig:matriz}
\end{figure}

\paragraph{Own band versus other bands.} For entities of each budget band we estimate bunching at the cutoff of each of the five bands. Figure~\ref{fig:matriz} shows the resulting $5\times5$ matrix (same universe and estimator, polynomial of degree 7): $b$ averages \IdMatrizDiagminima{} on the diagonal and \IdMatrizFueraminima{} off it at the small-purchase cutoff, and \IdMatrizDiagmenor{} versus \IdMatrizFueramenor{} at the tender cutoff. Entities bunch at their own threshold and essentially nowhere else; the off-diagonal cells are descriptive, since an entity's own spike can fall inside the fit region of a neighboring band's cutoff.

\begin{figure}[H]\centering
\includegraphics[width=\textwidth]{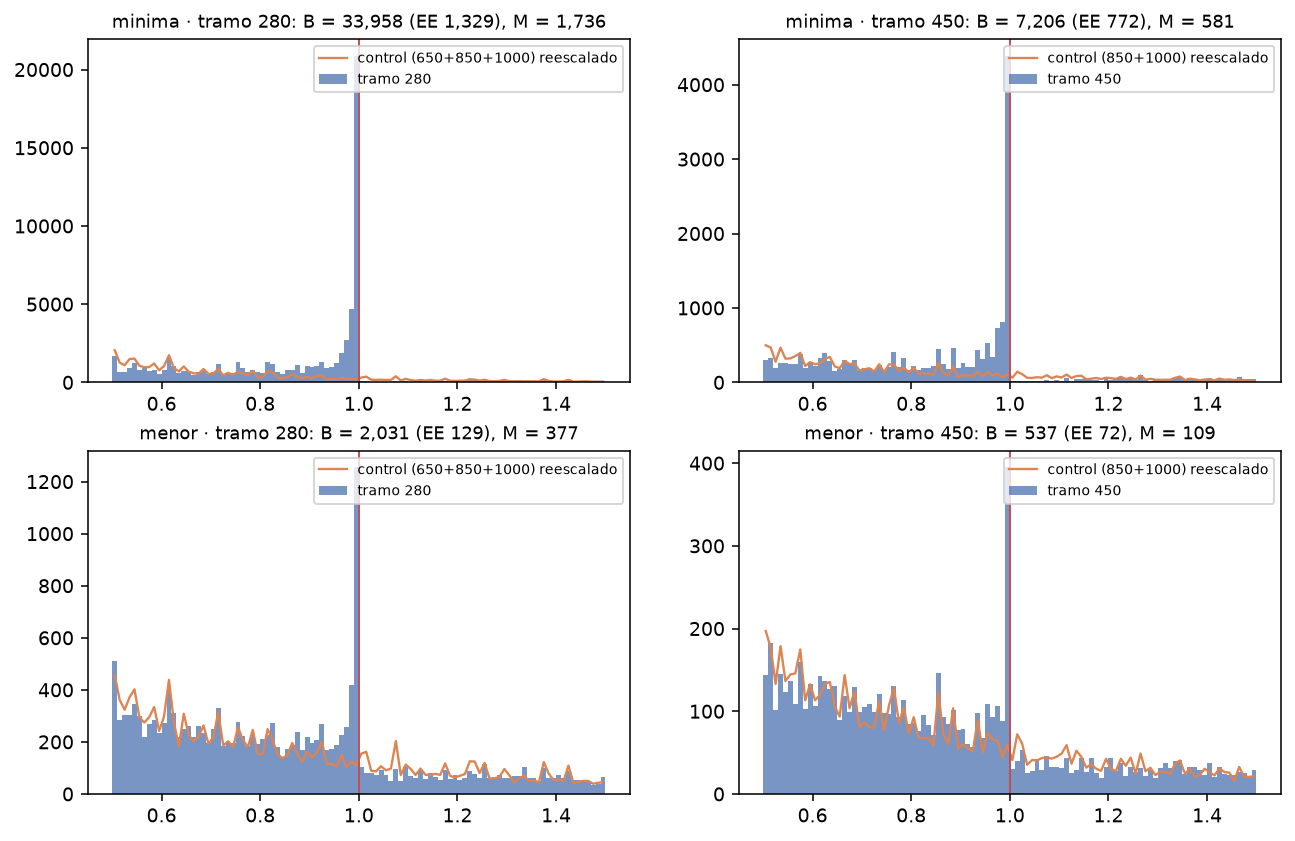}
\caption{Cross-band counterfactual. Blue bars: entities of the band in the panel title; orange line: density of entities of the control bands, in the same pesos and year, rescaled by a quadratic fit of the log ratio outside (0.90, 1.10].}\label{fig:dib}
\end{figure}

\paragraph{Cross-band counterfactual.} Using the density of entities from higher bands, in the same pesos and year, as the counterfactual (Section~\ref{sec:dib}) gives, for the 280 and 450 bands, $B = \Bdibminima$ (s.e.\ \BdibEEminima) at the small-purchase cutoff and \Bdibmenor{} (s.e.\ \BdibEEmenor) at the tender cutoff, against \BpolyDosTramosminima{} and \BpolyDosTramosmenor{} from the polynomial on the same two bands. The cross-band counterfactual is somewhat lower above the cutoff than the polynomial, and $B$ correspondingly larger, but the two designs agree on the order of magnitude and on the share of bin mass. Figure~\ref{fig:dib} shows the treated and rescaled control densities: the control tracks the round-number spikes of the treated density everywhere except at the treated threshold.

\subsection{Placebos}

Table~\ref{tab:placebos} applies the same estimator to samples that face no cutoff at the point examined. Consulting services must go through a merit contest at any value, so the tender threshold is irrelevant to them: $B = \SubObjetoConsultoriamenorB$ (s.e.\ \SubObjetoConsultoriamenorBEE), a strict zero. The other placebos are not zero but they are small: reverse auctions and merit contests together, direct contracting, and special-regime procedures show $b$ between 0.4 and 1.2, against 8 and 31 in the treated samples, with excess mass of 7 to 19 percent of the observed bin mass against 62 and 86 percent, and no missing mass above the cutoff. Their excess is concentrated in a one-bin spike at the exact value of the threshold: the threshold works as a focal number for budgeting even where it does not bind, and special-regime manuals often copy the statute's cutoffs. We therefore read the placebos as bounding the focal-number component of the treated estimates at roughly a tenth, not as showing that it is absent. In SECOP~I, which classifies each process as general-statute or special-regime, direct contracting and special-regime processes show $b = \PlSIDirectamenorb$ and $\PlSIRegimenEspecialmenorb$ at the tender cutoff. At the small-purchase cutoff the SECOP~I placebos are again small but not zero ($b = \PlSIDirectaminimab$ and $\PlSIRegimenEspecialminimab$).

\begin{table}[H]\centering\small
\caption{Placebo samples: procedures without a value cap at the cutoff examined}\label{tab:placebos}
\input{tablas/placebos.tex}
\end{table}

\subsection{Heterogeneity and robustness}\label{sec:robustez}

Table~\ref{tab:submuestras} re-estimates $B$ and $b$ by subsample. Three results stand out. First, territorial entities---municipalities and departments, governed by elected mayors and governors, together with their decentralized agencies---bunch more than national agencies run by appointed officials: $b = \SubOrdenTerritorialminimab$ against \SubOrdenNacionalminimab{} at the small-purchase cutoff and \SubOrdenTerritorialmenorb{} against \SubOrdenNacionalmenorb{} at the tender cutoff. In share terms the gap is smaller, \SubOrdenTerritorialminimaPart{} against \SubOrdenNacionalminimaPart{} of the observed bin mass, because the share saturates near one; in excess processes per process in the window it is \SubOrdenTerritorialminimaBN{} against \SubOrdenNacionalminimaBN. This is the opposite of the Italian pattern documented by \citet{Coviello2026}, where appointed officials manipulate more. Part of the gap is composition---territorial entities are concentrated in the lowest band, where bunching is strongest. Reweighting national processes to the territorial distribution of band $\times$ object $\times$ year raises their $b$ only to \RepondNacionalComoTerritorialminimab{} (s.e.\ \RepondNacionalComoTerritorialminimabEE) against \RepondTerritorialObjetivominimab{} for territorial entities at the small-purchase threshold; at the tender threshold the reweighted gap (\RepondNacionalComoTerritorialmenorb{} against \RepondTerritorialObjetivomenorb, s.e.\ \RepondNacionalComoTerritorialmenorbEE) is not statistically significant. The gap also appears within cells: in the 280 SMMLV band, $b = \SubOrdenTramoTerritorialTwoEightZerominimab$ for territorial against \SubOrdenTramoNacionalTwoEightZerominimab{} for national entities; in the 450 band, \SubOrdenTramoTerritorialFourFiveZerominimab{} against \SubOrdenTramoNacionalFourFiveZerominimab; for goods, \SubOrdenObjetoTerritorialBienesminimab{} against \SubOrdenObjetoNacionalBienesminimab; for services, \SubOrdenObjetoTerritorialServiciosminimab{} against \SubOrdenObjetoNacionalServiciosminimab; and for public works at the tender threshold, \SubOrdenObjetoTerritorialObramenorb{} against \SubOrdenObjetoNacionalObramenorb. The regional environmental authorities, whose directors are appointed but which sit within the territorial order, bunch like national agencies ($b = \SubOrdenCorporacionAutonomaminimab$). Second, the result does not depend on the self-declared threshold: on the subsample whose band comes only from budget data, $b = \SubUmbralIndependienteCuipoPgnminimab$ (s.e.\ \SubUmbralIndependienteCuipoPgnminimabEE) and \SubUmbralIndependienteCuipoPgnmenorb{} (s.e.\ \SubUmbralIndependienteCuipoPgnmenorbEE). Third, dropping estimated values that are exact multiples of one million pesos leaves $b$ at \SubRedondoNoMultiplominimab{} and \SubRedondoNoMultiplomenorb. Bunching is present in every band, every year from 2017 to 2026, and every object type except consulting; it is strongest for public works at the tender cutoff ($b = \SubObjetoObramenorb$) and weakest for goods ($\SubObjetoBienesmenorb$), which have the reverse auction as an alternative.

\begin{table}[H]\centering\small
\caption{Bunching by subsample}\label{tab:submuestras}
\begin{adjustbox}{max width=\textwidth}
\input{tablas/submuestras.tex}
\end{adjustbox}
\begin{minipage}{\textwidth}\footnotesize Left block: small-purchase threshold; right block: tender threshold. Entity-block bootstrap standard errors.\end{minipage}
\end{table}

\paragraph{Two further thresholds.} SECOP~I replicates the result for 2011--2020, before the transactional platform existed. The band of each entity is the modal band observed in SECOP~II applied backwards, which covers 78 percent of SECOP~I processes; for 2016--2019, where the entity's band in the same year is also observed, using it instead changes nothing ($b = \RobSIContemporaneominimab$ against \RobSIModalminimab{} with the modal band, on \RobSIN{} processes). With that threshold, $B = \SIBminima$ (s.e.\ \SIBEEminima), \SIPartminima{} of the mass below the small-purchase cutoff, present every year (Appendix Table~\ref{tab:secop1}). And public schools, whose own cutoff is 20 SMMLV under a separate regulation, bunch below it too: $B = \FseB$ (s.e.\ \FseBEE), $b = \Fseb$, across \FseN{} processes.

%% file: tablas/principal.tex
\begin{tabular}{lrrrrrrrr}
\toprule
 & Excess mass $B$ & (s.e.) & Share of bin mass & $b$ & (s.e.) & Missing mass $M$ & Exactly at threshold & $N$ in window \\
\midrule
Small-purchase threshold (10\% of MC) & 49,631 & (2,076) & 86\% & 30.8 & (0.7) & 3,536 & 12,561 & 167,349 \\
\quad excluding values exactly at the threshold & 37,070 & (1,524) & 82\% & 23.0 & (0.6) &  & 0 & 154,788 \\
\quad integration-constraint window ($z_U$ = 1.25) & 45,463 & (1,900) & 79\% & 18.6 & (0.6) & 29,759 &  &  \\
\quad cross-band counterfactual & 41,165 & (1,614) & 92\% &  &  &  &  & 94,138 \\
\midrule
Tender threshold (MC) & 3,157 & (182) & 63\% & 8.4 & (0.5) & 301 & 451 & 43,337 \\
\quad excluding values exactly at the threshold & 2,706 & (152) & 59\% & 7.2 & (0.4) &  & 0 & 42,886 \\
\quad integration-constraint window ($z_U$ = 1.23) & 2,760 & (170) & 55\% & 6.1 & (0.4) & 2,890 &  &  \\
\quad cross-band counterfactual & 2,569 & (147) & 57\% &  &  &  &  & 23,971 \\
\bottomrule
\end{tabular}

%% file: tablas/placebos.tex
\begin{tabular}{llrrrr}
\toprule
Sample & Threshold & Excess mass $B$ & (s.e.) & $b$ & (s.e.) \\
\midrule
\textit{SECOP II, 2015--2026} &  &  &  &  &  \\
Consulting (merit contest at any value) & Tender & -15 & (17) & -0.3 & (0.4) \\
Reverse auctions and merit contests (no value cap) & Tender & 153 & (45) & 1.0 & (0.3) \\
Direct contracting & Small purchase & 2,781 & (706) & 0.4 & (0.1) \\
Direct contracting & Tender & 248 & (63) & 0.7 & (0.2) \\
Special-regime procedures & Small purchase & 976 & (296) & 0.8 & (0.3) \\
Special-regime procedures & Tender & 447 & (124) & 1.2 & (0.3) \\
\midrule
\textit{SECOP I, 2011--2020} &  &  &  &  &  \\
Special regime (classified per process) & Small purchase & 2,005 & (453) & 2.3 & (0.5) \\
Direct contracting & Small purchase & 8,783 & (1,339) & 1.4 & (0.2) \\
Special regime (classified per process) & Tender & 108 & (75) & 0.4 & (0.4) \\
Direct contracting & Tender & 71 & (77) & 0.2 & (0.2) \\
\midrule
\textit{For comparison: treated sample} &  &  &  &  &  \\
All competitive procedures & Small purchase & 49,631 & (2,076) & 30.8 & (0.7) \\
All competitive procedures & Tender & 3,157 & (182) & 8.4 & (0.5) \\
\bottomrule
\end{tabular}

%% file: tablas/submuestras.tex
\begin{tabular}{lrrrrrrrr}
\toprule
 & $B$ & (s.e.) & $b$ & (s.e.) & $B$ & (s.e.) & $b$ & (s.e.) \\
\midrule
\textit{Level of government} &  &  &  &  &  &  &  &  \\
Territorial (elected) & 35,595 & (1,449) & 37.3 & (1.0) & 2,243 & (157) & 10.2 & (0.8) \\
National (appointed) & 13,594 & (1,284) & 22.3 & (1.1) & 900 & (107) & 6.3 & (0.7) \\
Regional environmental authorities & 573 & (142) & 24.3 & (2.4) & 37 & (24) & 3.9 & (2.4) \\
\textit{Object} &  &  &  &  &  &  &  &  \\
Public works & 3,632 & (333) & 39.0 & (2.4) & 1,208 & (83) & 15.7 & (1.4) \\
Goods & 22,318 & (1,040) & 31.7 & (0.9) & 531 & (48) & 4.6 & (0.5) \\
Services & 19,207 & (708) & 29.3 & (0.9) & 1,161 & (85) & 9.2 & (0.7) \\
Consulting & 2,377 & (204) & 26.2 & (2.5) & -15 & (17) & -0.3 & (0.4) \\
\textit{Threshold source} &  &  &  &  &  &  &  &  \\
Self-declared (PAA) or imputed & 47,754 & (2,061) & 31.5 & (0.8) & 3,103 & (149) & 8.4 & (0.5) \\
Budget-based (CUIPO/PGN) & 1,878 & (146) & 19.0 & (1.6) & 50 & (14) & 5.7 & (1.9) \\
\textit{Round numbers} &  &  &  &  &  &  &  &  \\
Not a multiple of COP 1M & 35,741 & (1,477) & 31.6 & (0.9) & 2,295 & (132) & 7.7 & (0.4) \\
Multiple of COP 1M & 14,231 & (809) & 34.3 & (1.1) & 858 & (72) & 10.8 & (1.2) \\
\textit{Budget band (SMMLV)} &  &  &  &  &  &  &  &  \\
280 & 28,435 & (1,269) & 48.7 & (1.4) & 1,611 & (113) & 11.0 & (0.7) \\
450 & 5,788 & (743) & 28.4 & (1.6) & 483 & (60) & 7.8 & (0.9) \\
650 & 3,500 & (624) & 24.2 & (1.9) & 194 & (45) & 5.0 & (1.1) \\
850 & 1,921 & (354) & 34.3 & (2.8) & 135 & (43) & 8.0 & (2.1) \\
1000 & 10,604 & (1,137) & 21.1 & (1.1) & 771 & (114) & 7.4 & (0.7) \\
\textit{Presidential-election ban} &  &  &  &  &  &  &  &  \\
Published during the ban & 6,975 & (270) & 37.2 & (1.7) & 399 & (34) & 8.6 & (0.8) \\
Outside the ban & 42,753 & (2,010) & 30.4 & (0.7) & 2,757 & (174) & 8.4 & (0.6) \\
\bottomrule
\end{tabular}

%% file: sec_mecanismos.tex
\section{Where the excess comes from, and what it costs}\label{sec:mecanismos}

\subsection{Missing mass}

If buyers only shaved values that would otherwise land slightly above the cutoff, the missing mass above it would equal the excess below. It does not. With the cross-band counterfactual, which extrapolates less than a polynomial, the mass missing between the cutoff and 1.5 times the cutoff is \MasaminimaHOneFive{} processes at the small-purchase threshold (\MasaPctminimaHOneFive{} of $B$) and \MasamenorHOneFive{} at the tender threshold (\MasaPctmenorHOneFive{} of $B$); Table~\ref{tab:masa} gives the profile. The rest of the excess at the small-purchase threshold does not come from a moderately larger project that was cut down. The gap is larger for territorial entities (\MasaPctTerritorialminimaHOneFive{} of the excess reappears up to 1.5 times the cutoff at the small-purchase threshold, \MasaPctTerritorialmenorHOneFive{} at the tender threshold) than for national ones (\MasaPctNacionalminimaHOneFive{} and \MasaPctNacionalmenorHOneFive), but even for national entities we do not find the near-complete matching of excess and missing mass that \citet{IvarsOlave2025} report; their sample (2018--2024, dominated by reverse auctions) and their window differ from ours, and reconciling the two estimates is a task for the next version. Two other margins can supply the remainder: splitting a larger purchase into several, and shifting purchases that would have gone through another route (direct contracting, framework agreements) into the small-purchase procedure because it is cheap. Our data can test the first directly.

\begin{table}[H]\centering\small
\caption{Missing mass above the threshold, cumulative}\label{tab:masa}
\input{tablas/masa.tex}
\begin{minipage}{\textwidth}\footnotesize Cross-band counterfactual, bands 280, 850 and 1000 against their controls; log ratio fitted with a linear term outside (0.90, 1.50]. In parentheses, share of the excess below the cutoff.\end{minipage}
\end{table}

\subsection{Splitting}

For every small purchase we count its ``siblings'': processes of the same entity, in the same UNSPSC product family, published within 45 days. Bunched small purchases (those in the last five percent below the cutoff) are not more likely to have a sibling than small purchases far from the cutoff (\FraccTieneHermDif, s.e.\ \FraccTieneHermEE, with entity-year fixed effects), but they are more likely to have a sibling awarded to the same supplier (\FraccHermMismoProvDif, s.e.\ \FraccHermMismoProvEE, on a base of \FraccHermMismoProvLejos). Whether the siblings' values sum above the threshold (\FraccHermSumaSobreUmbralDif) is not informative, since a bunched process is by construction within five percent of it. With this measure, splitting is present but explains little of the excess.

\subsection{Post-award additions}

Colombian law allows contract additions of up to fifty percent of the initial value. We measure additions as the current value of the contract over the awarded value, validated against the modification records (the value grows by more than one percent in 31 percent of contracts with a recorded modification and in 2 percent of those without). Two facts stand out in Table~\ref{tab:consecuencias}. Contracts that were bunched just below the tender threshold are more likely to receive an addition than contracts awarded just above it (\ResPAdicionValormenorAm{} against \ResPAdicionValormenorArr, a difference of \ResPAdicionValormenorDif, s.e.\ \ResPAdicionValormenorDifEE) and than non-manipulated processes of the same value (\ResPAdicionValormenorDifNM, s.e.\ \ResPAdicionValormenorDifNMEE), and more likely to reach the legal cap of fifty percent (\ResPAdicionTopemenorAm{} against \ResPAdicionTopemenorArr). Since an addition of more than a few percent takes a bunched contract across the threshold it avoided, the threshold is crossed after the award, once the procedure has been chosen. These outcomes condition on awarded processes with a single contract.

\subsection{Competition and awardees}

Table~\ref{tab:consecuencias} reports, for each outcome, the mean among bunched processes recovered with the decomposition of Section~\ref{sec:dp}, the mean just above the cutoff, and the mean among non-manipulated processes of the same value. The ``just above'' comparison is the policy-relevant one---what the same purchases would have faced under the more demanding procedure---while the ``same value'' comparison isolates selection into bunching. At the tender threshold the alternative to an abbreviated selection is the open tender, so ``just above'' uses abbreviated selections and tenders only; reverse auctions and merit contests, which are chosen by object and carry no cap, are excluded from that comparison. At the small-purchase threshold the alternative depends on the object, and ``just above'' pools the three procedures. At the tender threshold, bunched processes attract \ResNProponentesmenorAm{} bidders against \ResNProponentesmenorArr{} for processes just above it (a difference of \ResNProponentesmenorDif, s.e.\ \ResNProponentesmenorDifEE) but only \ResNProponentesmenorDifNM{} fewer (s.e.\ \ResNProponentesmenorDifNMEE) than non-manipulated processes of the same value; \ResUnSoloProponentemenorAm{} of them receive a single bid against \ResUnSoloProponentemenorArr; and the discount obtained on the estimated value is \ResDescuentomenorAm{} against \ResDescuentomenorArr{} (\ResDescuentomenorDif, s.e.\ \ResDescuentomenorDifEE), a gap that is also present against non-manipulated processes (\ResDescuentomenorDifNM). The discount computed from contract values rather than the award record, which does not condition on the award flag, shows the same gap (\ResDescuentoContratomenorDif). At the small-purchase threshold the number of bidders is barely affected (\ResNProponentesminimaDif{} against processes just above), but who wins changes: bunched processes go to a recurrent supplier of the entity in \ResRecurrenteminimaAm{} of cases against \ResRecurrenteminimaArr{} just above the cutoff and \ResRecurrenteminimaNoManip{} among non-manipulated processes of the same value, and to an individual rather than a firm in \ResPersonaNaturalminimaAm{} against \ResPersonaNaturalminimaArr; part of the latter reflects that registration in the suppliers' registry is not required below the small-purchase threshold. There is no difference in the share of awards to micro, small and medium firms. Outcomes are missing for some processes (no bidder list, no contract); the decomposition assumes they are missing at random within a bin.

\begin{table}[H]\centering\small
\caption{Outcomes of bunched processes}\label{tab:consecuencias}
\begin{adjustbox}{max width=\textwidth}
\input{tablas/consecuencias.tex}
\end{adjustbox}
\begin{minipage}{\textwidth}\footnotesize ``Bunchers'': mean outcome of the excess mass, recovered as in Section~\ref{sec:dp}. ``Just above'': linear extrapolation of bin means from (1.04, 1.45] to the cutoff. ``Non-manip.'': quadratic extrapolation from (0.60, 0.95]. Entity-block bootstrap standard errors (200 replications). Bidders and single-bidder status come from the list of bidders registered per process, not from the number of responses field.\end{minipage}
\end{table}

\begin{figure}[H]\centering
\includegraphics[width=\textwidth]{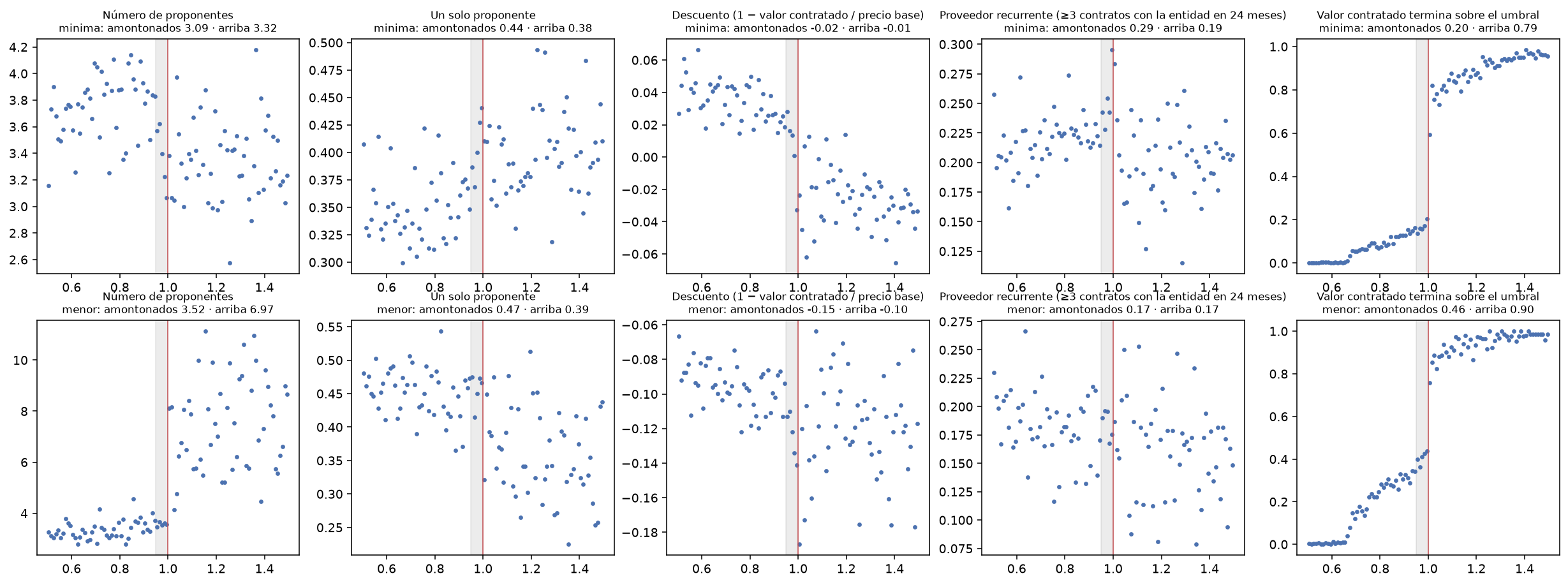}
\caption{Mean outcomes by bin of the normalized estimated value. Top row: small-purchase threshold; bottom row: tender threshold. Shaded: the bunching window.}\label{fig:resultados}
\end{figure}

\subsection{Political heterogeneity}\label{sec:politica}

\input{sec_politica.tex}

%% file: tablas/masa.tex
\begin{tabular}{lrrrrr}
\toprule
Threshold & up to 1.04 & up to 1.10 & up to 1.25 & up to 1.50 & Excess $B$ \\
\midrule
Small-purchase threshold (10\% of MC) & 1,913 (5\%) & 4,259 (11\%) & 8,096 (20\%) & 12,327 (31\%) & 39,691 \\
Tender threshold (MC) & 302 (13\%) & 712 (30\%) & 1,370 (57\%) & 1,922 (80\%) & 2,390 \\
\bottomrule
\end{tabular}

%% file: tablas/consecuencias.tex
\begin{tabular}{lrrrrrr}
\toprule
Outcome & Bunchers & Just above & Diff. & (s.e.) & Diff. vs non-manip. & (s.e.) \\
\midrule
\textit{Tender threshold (MC)} &  &  &  &  &  &  \\
Number of bidders & 3.52 & 6.97 & -3.45 & (1.02) & -0.20 & (0.53) \\
Single bidder & 47.3\% & 39.1\% & +8.1 & (2.7) & +3.8 & (3.2) \\
Discount on estimated value (awarded processes) & 4.6\% & 8.7\% & -4.2 & (1.1) & -0.8 & (1.1) \\
Discount on estimated value (contract value, all) & -14.9\% & -10.5\% & -4.4 & (1.3) & -5.4 & (1.4) \\
No contract registered & 16.4\% & 18.1\% & -1.7 & (1.8) & -1.7 & (2.2) \\
Awarded to an individual & 11.5\% & 9.8\% & +1.7 & (1.7) & +2.3 & (2.1) \\
Recurrent supplier (3+ contracts, 24 months) & 17.5\% & 16.8\% & +0.7 & (2.1) & -1.5 & (2.6) \\
Contract value grows > 1\% over award & 48.9\% & 44.1\% & +4.8 & (3.1) & +9.8 & (3.2) \\
Addition at the 50\% legal cap & 16.2\% & 10.9\% & +5.3 & (2.1) & +4.8 & (2.3) \\
Any contract modification (2017--2024) & 79.9\% & 82.1\% & -2.1 & (2.2) & -2.6 & (3.0) \\
\textit{Small-purchase threshold (10\% of MC)} &  &  &  &  &  &  \\
Number of bidders & 3.09 & 3.32 & -0.23 & (0.14) & -0.62 & (0.09) \\
Single bidder & 43.7\% & 38.2\% & +5.5 & (1.8) & +6.4 & (1.1) \\
Discount on estimated value (awarded processes) & 7.4\% & 9.0\% & -1.6 & (0.7) & -2.5 & (0.4) \\
Discount on estimated value (contract value, all) & -2.4\% & -0.8\% & -1.6 & (0.8) & -3.9 & (0.6) \\
No contract registered & 15.1\% & 24.2\% & -9.0 & (1.0) & -0.1 & (0.7) \\
Awarded to an individual & 24.4\% & 12.8\% & +11.5 & (1.1) & +4.6 & (0.9) \\
Recurrent supplier (3+ contracts, 24 months) & 28.7\% & 19.3\% & +9.4 & (1.8) & +6.4 & (1.0) \\
Contract value grows > 1\% over award & 24.4\% & 25.6\% & -1.2 & (1.6) & +0.3 & (1.1) \\
Addition at the 50\% legal cap & 10.9\% & 6.1\% & +4.8 & (0.9) & +3.7 & (0.8) \\
Any contract modification (2017--2024) & 65.9\% & 76.6\% & -10.7 & (1.8) & -2.7 & (1.3) \\
\bottomrule
\end{tabular}

%% file: sec_politica.tex
Municipal governments are the most intense bunchers in the data: among mayoral administrations (entities named ``Municipio de'' or ``Alcald\'ia'') the excess below the small-purchase threshold is \PolMargenminimaTOneb{} to \PolMargenminimaTThreeb{} times the counterfactual bin. To ask whether electoral competition disciplines them, we link each municipality to the mesa-level results of the 2019 and 2023 mayoral elections and split administrations into terciles of the winner's margin of victory (the 2019 election governs processes of 2020--2023; the 2023 election those of 2024--2026). We can match \PolMunicipios{} municipalities by name, covering \PolCruzados{} of \PolProcesosAlcaldias{} municipal processes.

The margin of victory does not predict bunching at the small-purchase threshold: $b = \PolMargenminimaTOneb$ (s.e.\ \PolMargenminimaTOnebEE) in the tercile of closest elections, \PolMargenminimaTTwob{} in the middle tercile and \PolMargenminimaTThreeb{} in the tercile of most lopsided ones. At the tender threshold the safest administrations bunch somewhat less (\PolMargenmenorTThreeb{} against \PolMargenmenorTOneb{} and \PolMargenmenorTTwob), a difference of about one standard error. Administrations elected by less than five points show $b = \PolMargenminimaCerradaFiveppb$ and \PolMargenmenorCerradaFiveppb; the rest, \PolMargenminimaNoCerradab{} and \PolMargenmenorNoCerradab. Whatever disciplines threshold manipulation in Colombian municipalities, it is not the closeness of the last election.

We also link the 2019 campaign-finance records of mayoral candidates to the suppliers awarded by the elected mayor's administration in 2020--2023. Matched donors are rare among awardees at any distance from the threshold---\PolDonmenorAmontonado{} of contracts bunched below the tender threshold against \PolDonmenorLejos{} far from it, on a few thousand contracts---so the data cannot speak to whether bunched processes reward donors; \citet{Gulzar2025}, with the full donor panel, find that they do.

%% file: sec_garantias.tex
\section{Electoral bans: the timing margin}\label{sec:garantias}

The four-month ban on direct contracting before presidential elections is the second source of variation in discretion. Colombian municipalities display a political budget cycle in the composition of spending \citep{DrazenEslava2010}; the only study of the ban itself, \citet{Canon2022}, finds no effect in a first-difference specification on monthly aggregates that removes the level shift we document here. Unlike the thresholds, it does not change what a buyer may buy but when and how: direct contracting is prohibited, the competitive procedures remain open, and the exempted sectors (health, defense, public credit, emergencies) keep contracting directly.

\begin{figure}[H]\centering
\includegraphics[width=\textwidth]{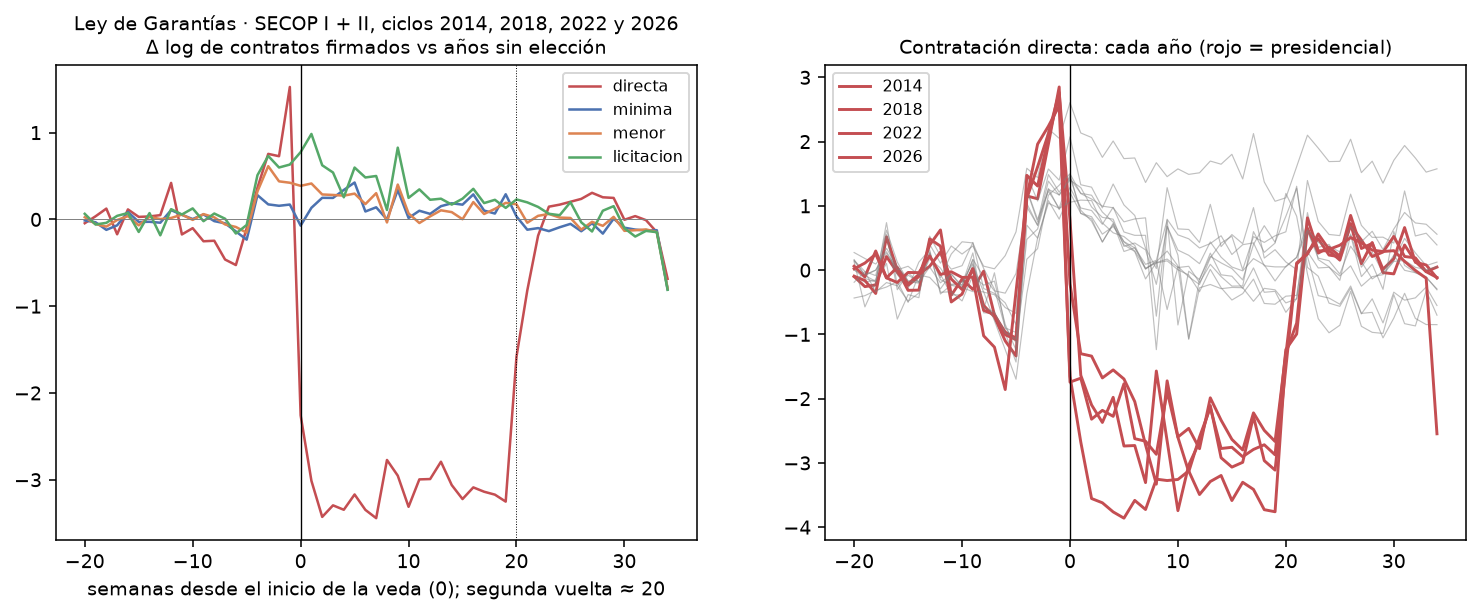}
\caption{Contracts signed by week around the start of the presidential-election ban on direct contracting. Left: difference in log counts between presidential and non-election years, normalized by weeks $-20$ to $-9$; SECOP~I and SECOP~II pooled, 2012--2026 excluding 2020. Right: direct contracting, each year separately (red: presidential years).}\label{fig:garantias}
\end{figure}

Figure~\ref{fig:garantias} plots, for each procedure, the difference in log weekly contracts between the four presidential years (2014, 2018, 2022, 2026) and the ten non-election years, in weeks relative to the start of the ban, pooling SECOP~I and SECOP~II. Table~\ref{tab:garantias} summarizes it. Direct contracting changes by \LGDirectaVeda{} during the ban ($p = \LGDirectaVedaP$, the smallest value attainable by exact randomization inference over the 1,001 assignments), and the drop has the same sign and a similar size in each of the four cycles ($\LGdirectaVedaTwoZeroOneFour$, $\LGdirectaVedaTwoZeroOneEight$, $\LGdirectaVedaTwoZeroTwoTwo$ and $\LGdirectaVedaTwoZeroTwoSix$ log points). A drop of this size mostly measures compliance with a legal prohibition; the informative margins are what happens before and after it. Buyers anticipate: in the two weeks before the ban direct contracting is \LGDirectaUltTwoEnero{} above its level in non-election years ($p = \LGDirectaUltTwoEneroP$), although the eight-week anticipation window as a whole is only \LGDirectaAnticipacion{} higher ($p = \LGDirectaAnticipacionP$). There is no rebound after the second round (\LGDirectaRebote, $p = \LGDirectaReboteP$). The net change in direct contracts from eight weeks before the ban to ten weeks after the election is \LGDirectaNeto{} ($p = \LGDirectaNetoP$): the point estimate suggests that anticipation does not fully compensate the ban, but we cannot reject that the ban only moves contracts in time.

\begin{table}[H]\centering\small
\caption{Electoral ban on direct contracting: contracts signed relative to non-election years}\label{tab:garantias}
\begin{adjustbox}{max width=\textwidth}
\input{tablas/garantias.tex}
\end{adjustbox}
\begin{minipage}{\textwidth}\footnotesize Each cell is $\exp(\bar\beta)-1$, where $\bar\beta$ is the mean over the indicated weeks of the difference in normalized log weekly counts between presidential years (2014, 2018, 2022, 2026) and the ten non-election years 2012--2026 (excluding 2020). In brackets, exact randomization-inference $p$-values over the 1,001 assignments of four treated years. Week 0 is the start of the ban; the second round falls in week 20.\end{minipage}
\end{table}

The ban shifts contracting toward the procedures it leaves open, and in particular toward the cheapest one: small purchases rise \LGMinimaVeda{} during the ban ($p = \LGMinimaVedaP$), abbreviated selections \LGMenorVeda{} and open tenders \LGLicitacionVeda{}, the last two imprecisely. The exemptions matter: in SECOP~II, where entities can be classified, direct contracting by health and defense entities falls \LGIIDirectaExceptuadasVeda{} ($p = \LGIIDirectaExceptuadasVedaP$) against \LGIIDirectaNoExceptuadasVeda{} for the rest ($p = \LGIIDirectaNoExceptuadasVedaP$). Special-regime entities, which the law also covers, cut their contracting by \LGRegimenEspecialVeda{}.

Does the ban interact with the thresholds? Bunching below the small-purchase threshold is somewhat more intense for processes published during the ban ($b = \SubVedaArtThreeThreeDentrominimab$, s.e.\ \SubVedaArtThreeThreeDentrominimabEE) than outside it (\SubVedaArtThreeThreeFueraminimab; Table~\ref{tab:submuestras}), and unchanged at the tender threshold: the increase in small purchases during the ban adds bunched processes at least in proportion. Two further episodes give null results and are reported in Appendix~\ref{app:lg}: the milder restriction on intergovernmental agreements before local elections (article~38), and the eleven weeks of 2021--22 during which a budget law lifted that restriction before the Constitutional Court struck it down retroactively.

%% file: tablas/garantias.tex
\begin{tabular}{lrrrrrrrrrr}
\toprule
Procedure & Last 2 weeks & $p$ & Weeks $-8$ to $-1$ & $p$ & Ban (weeks 0--16) & $p$ & Weeks 21--30 & $p$ & Net ($-8$ to $+30$) & $p$ \\
\midrule
Direct contracting & +209\% & [0.001] & +29\% & [0.181] & -95\% & [0.001] & +6\% & [0.872] & -26\% & [0.177] \\
Small purchases & +18\% & [0.232] & +4\% & [0.665] & +18\% & [0.005] & -9\% & [0.126] & +6\% & [0.239] \\
Abbreviated selection & +54\% & [0.308] & +21\% & [0.504] & +21\% & [0.329] & -2\% & [0.888] & +6\% & [0.756] \\
Open tenders & +85\% & [0.117] & +34\% & [0.281] & +56\% & [0.223] & +7\% & [0.844] & +21\% & [0.493] \\
Special-regime procedures & +66\% & [0.166] & +2\% & [0.953] & -46\% & [0.001] & -28\% & [0.292] & -24\% & [0.187] \\
\bottomrule
\end{tabular}

%% file: sec_chile.tex
\section{Chile: fixed thresholds and a cutoff that disappeared}\label{sec:chile}

Chile provides a comparison with thresholds that do not depend on the buyer and move only with the monthly indexation of the tax unit. We apply the same estimator to the universe of tenders published on Mercado P\'ublico between 2015 and 2026, one record per tender, with the estimated amount converted to UTM at the month of publication. The preferred sample keeps tenders whose estimated amount was visible to bidders, since a hidden amount cannot coordinate bids; all tenders are reported as a robustness check. Standard errors resample buying organizations.

Table~\ref{tab:chile} shows bunching below the 100 UTM cutoff (L1 to LE), $B = \ChlUOneZeroZeroVisibleB$ (s.e.\ \ChlUOneZeroZeroVisibleBEE), $b = \ChlUOneZeroZeroVisibleb$, and below the 1,000 UTM cutoff (LE to LP), $B = \ChlUOneZeroZeroZeroVisibleB$, $b = \ChlUOneZeroZeroZeroVisibleb$; a smaller spike at 5,000 UTM; and a weak one at 2,000 UTM while that cutoff existed. Including tenders with hidden amounts doubles or triples $B$ and leaves $b$ essentially unchanged. The excess is stable year by year from 2015 to 2026 at both main cutoffs (Appendix Table~\ref{tab:chile_anio}).

\begin{figure}[H]\centering
\includegraphics[width=0.85\textwidth]{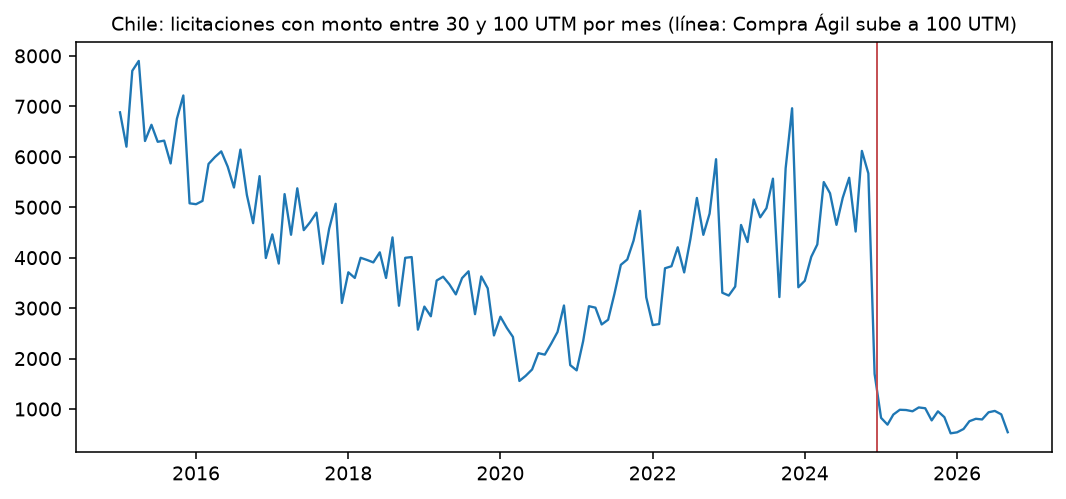}
\caption{Chile: monthly number of tenders with estimated amounts between 30 and 100 UTM. The vertical line marks the entry into force of Law 21.634, which raised the Compra \'Agil ceiling to 100 UTM.}\label{fig:chile}
\end{figure}

Two features of the Chilean setting complement the Colombian evidence. First, the 2024 regulation abolished the 2,000 UTM cutoff: in 2025--26 the estimator finds $B = \ChlUTwoZeroZeroZeroPlaceboTwoZeroTwoFiveTwoZeroTwoSixB$ (s.e.\ \ChlUTwoZeroZeroZeroPlaceboTwoZeroTwoFiveTwoZeroTwoSixBEE) at a point where a cutoff used to be. Second, Law 21.634 raised the Compra \'Agil ceiling from 30 to 100 UTM on 12 December 2024: the monthly number of tenders with estimated amounts between 30 and 100 UTM fell from \ChlCApre{} to \ChlCApost{} (\ChlCAcaida), as buyers moved those purchases to the direct channel the reform opened (Figure~\ref{fig:chile}).

\begin{table}[H]\centering\small
\caption{Chile: bunching below the UTM thresholds, 2015--2026}\label{tab:chile}
\begin{adjustbox}{max width=\textwidth}
\input{tablas/chile.tex}
\end{adjustbox}
\begin{minipage}{\textwidth}\footnotesize Tenders in Chilean pesos with a positive estimated amount, converted to UTM at the month of publication. The 2,000 UTM cutoff existed until 2024; the last row estimates bunching at that point after its abolition. Organization-block bootstrap standard errors.\end{minipage}
\end{table}

\begin{figure}[H]\centering
\includegraphics[width=\textwidth]{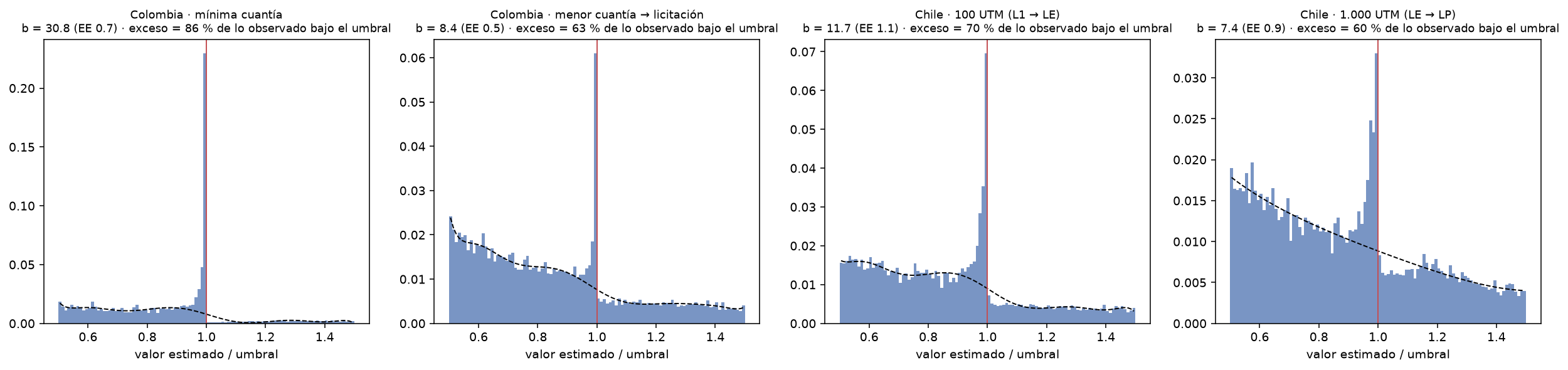}
\caption{Colombia and Chile with the same estimator. Density of estimated values relative to the cutoff; Colombia: all competitive procedures, SECOP~II 2015--2026; Chile: tenders with visible amounts, 2015--2026.}\label{fig:colombia_chile}
\end{figure}

Figure~\ref{fig:colombia_chile} puts the two countries side by side. At the lower cutoff Colombia bunches about three times more than Chile ($b = \bminima$ against \ChlUOneZeroZeroVisibleb); at the upper cutoff the two are similar (\bmenor{} against \ChlUOneZeroZeroZeroVisibleb). Two differences in the rules are candidates for the gap at the lower cutoff. In Colombia the small-purchase procedure removes almost every requirement, whereas the L1 tender in Chile is still a tender; and Colombia's cutoff is specific to the entity and indexed annually, so a buyer who tracks it can plan the year's purchases around it. Testing these explanations requires cross-country variation in the rules themselves, which the 2024 Chilean reform and the pending Colombian procurement reform will provide.

%% file: tablas/chile.tex
\begin{tabular}{llrrrrrr}
\toprule
Threshold & Sample & $B$ & (s.e.) & Share of bin mass & $b$ & (s.e.) & $N$ in window \\
\midrule
100 UTM & amount visible to bidders & 27,150 & (3,822) & 70\% & 11.7 & (1.0) & 228,924 \\
 & all & 65,503 & (10,092) & 69\% & 11.0 & (1.5) & 478,107 \\
1,000 UTM & amount visible to bidders & 5,969 & (948) & 60\% & 7.4 & (1.0) & 88,275 \\
 & all & 28,300 & (4,298) & 71\% & 12.3 & (1.4) & 219,743 \\
2,000 UTM & amount visible to bidders & 1,017 & (150) & 32\% & 2.4 & (0.4) & 48,339 \\
 & all & 4,340 & (544) & 47\% & 4.5 & (0.5) & 109,280 \\
 & 2025--26, cutoff abolished & -49 & (31) & -10\% & -0.5 & (0.3) & 11,044 \\
5,000 UTM & amount visible to bidders & 2,660 & (468) & 57\% & 6.7 & (0.9) & 37,216 \\
 & all & 12,715 & (1,926) & 71\% & 12.2 & (1.3) & 95,770 \\
\bottomrule
\end{tabular}

%% file: sec_conclusion.tex
\section{Discussion and conclusion}\label{sec:conclusion}

Colombian public buyers set the estimated value of what they buy just below the legal thresholds that would force a more demanding procedure. The excess is large---\Partminima{} of the processes observed in the last five percent below the small-purchase cutoff and \Partmenor{} below the tender cutoff are there because of the cutoff---and it is identified by variation that few settings offer: the cutoff differs across five budget bands and moves every January with the minimum wage, so we can show that the spike follows the threshold and not the round numbers, build a counterfactual from entities of other bands, and obtain zeros where the rules impose no cutoff. Elected local governments manipulate twice as much as appointed national agencies; bunched processes attract fewer bidders, obtain smaller discounts, go to recurrent suppliers and individuals, and are more often enlarged after award to end above the threshold they avoided. The four-month electoral ban on direct contracting shows the same buyers responding on the timing margin, front-loading direct contracts into the two weeks before the ban and shifting toward small purchases during it.

\paragraph{Limitations.} Four deserve emphasis. First, 93 percent of the estimation sample uses the threshold that the entity itself declares in its annual procurement plan, validated against the legal bands; the budget-based subsample, where the threshold is independent of the entity, gives the same sign and order of magnitude but is small, and extending it back to 2016 requires archival budget data. Second, the placebos are an order of magnitude below the treated samples but not always zero at the small-purchase cutoff: the threshold is a focal number for budgeting even where it does not bind, and in SECOP~I the modal band is an assumption. Third, the missing mass above the cutoffs accounts for \MasaPctminimaHOneFive{} of the excess at the small-purchase threshold within 1.5 times the cutoff; the remainder comes from margins---splitting, shifting purchases from other routes---that our data only partly identify, as \citet{Carril2026} and \citet{IvarsOlave2025} also find. Fourth, inference: standard errors that allow for band-year shocks are about five times larger than entity-clustered ones, and with four electoral cycles the electoral-ban results rest on randomization inference over years. Fifth, the outcome comparisons are descriptive: the decomposition recovers the mean outcome of the excess mass, but bunchers are selected, and the bounds of \citet{DiamondPersson2016} and \citet{Szucs2024} that would turn the differences into effects are left for the next version.

\paragraph{Policy.} Three implications follow. A threshold indexed to the buyer's budget and to the minimum wage is not harder to manipulate than a fixed one; if anything, buyers track it closely. The cost of manipulation is concentrated at the tender threshold, where competition halves, and in who wins at the small-purchase threshold. And rules that close one door---the electoral ban---move activity to the door left open, which is the procedure whose threshold is manipulated; a reform that raises the small-purchase threshold, as Colombia has debated and Chile did in 2024, expands the room for both.

\paragraph{Next steps.} The immediate extensions are the political heterogeneity (margin of victory and campaign donors, Section~\ref{sec:politica}), the historical budgets that would make the threshold independent of the entity throughout, a matched comparison of elected and appointed buyers by size and object, and, with co-authors in Chile, the purchase-order data that identify the Compra \'Agil channel directly.

%% file: sec_declaraciones.tex
\section*{Declarations}
\addcontentsline{toc}{section}{Declarations}

\paragraph{Author contributions (CRediT).} Cristian Espinal Maya: conceptualization, data curation, formal analysis, investigation, methodology, software, validation, visualization, writing -- original draft. Santiago Jim\'enez Londo\~no: writing -- review \& editing.

\paragraph{Data and code availability.} All data used in this paper are public: the SECOP~I and SECOP~II tables published by Colombia Compra Eficiente on Colombia's open-data portal (datos.gov.co), the territorial budget reports (CUIPO) and the national budget (PGN), mesa-level electoral results from the Registrar's office, campaign-finance reports (Cuentas Claras), and Chile's Mercado P\'ublico open-data archive. A replication sample (SECOP~II competitive processes published in 2023--2024, with each process's normalized estimated value) and the code that reproduces the headline bunching estimates are distributed as ancillary files with the arXiv version of this paper (folder \texttt{anc/}). The full analysis files and the remaining code are available from the corresponding author upon reasonable request.

\paragraph{Funding.} This research received no specific funding.

\paragraph{Competing interests.} Espinal is a co-founder of and partner in firms that provide consulting and technology services to Colombian public entities; the paper uses exclusively public open data and no entity is singled out. Jim\'enez Londo\~no's past professional work includes collaborations with Colombian public institutions, none of which is singled out in this paper; he declares no financial interest related to this work.

\paragraph{Use of AI tools.} AI-based assistants were used to help write and review code and to edit the manuscript. The authors reviewed all output, take full responsibility for the content, and every reported number is recomputed from the data by the verification battery described in Appendix~\ref{app:datos}.

%% file: apendice.tex
\section{Data construction and verification}\label{app:datos}

\paragraph{Download.} Every table was downloaded from the national open-data portal (Socrata API) in pages of 25,000 rows ordered by the row identifier, partitioned by year of the table's reference date. Each page was checked against the number of rows expected from the server's own count before being accepted; pages with an embedded server error or a row-count mismatch were re-requested. The raw CSV of every page and a typed Parquet copy are archived with SHA-256 hashes. A reconciliation table records, for every table and partition, the rows expected, downloaded, kept and excluded, and the count of distinct row identifiers.

\paragraph{Cleaning.} No row is dropped on content; rows are excluded only when the key is missing, and the count of exclusions is recorded by rule. Tax identifiers are normalized by stripping the check digit; suppliers without a numeric identifier (consortia) are keyed by their normalized name. Processes are aggregated to the portfolio identifier (Section~\ref{sec:datos}), keeping the record of the most advanced stage and the estimated value of the earliest one. Contracts are keyed by contract identifier and linked to processes through the portfolio identifier; bidders are linked through the stage identifier at which offers are submitted.

\paragraph{Thresholds.} Section~\ref{sec:umbrales}. The band from the annual procurement plan is accepted only within three percent of a legal band; the rule also detects swapped fields. The budget-based band uses the initial appropriation of the expenditure budget reported in the first quarter (territorial entities, 2023 on) or in January (national entities, 2019 on). Where both exist they agree in 92 percent of entity-years.

\paragraph{Verification.} Every number in the text is produced by an analysis script and written to a claims registry with its provenance. A separate battery recomputes the headline numbers with independently written queries---the bin counts, the sample sizes, the excess mass, the electoral-ban statistics and the Chilean estimates---and reconciles every table and partition; the current run performs 512 checks with no discrepancy.

\section{Robustness}\label{app:robustez}

\begin{table}[H]\centering\small
\caption{Excess mass $B$ across specifications}\label{tab:robustez}
\input{tablas/robustez.tex}
\begin{minipage}{\textwidth}\footnotesize Each cell is $B$ for the polynomial degree and lower limit $z_L$ in the row and the upper limit $z_U$ in the column; the last column widens $z_U$ until the missing mass equals the excess (capped at 1.25).\end{minipage}
\end{table}

\begin{table}[H]\centering\small
\caption{SECOP I, 2011--2020: bunching by year}\label{tab:secop1}
\input{tablas/secop1.tex}
\begin{minipage}{\textwidth}\footnotesize Threshold: modal band of the entity in SECOP~II, applied to the SMMLV of each year. Placebo rows use the special-regime and direct-contracting classification recorded per process in SECOP~I.\end{minipage}
\end{table}

\begin{table}[H]\centering\small
\caption{Chile: bunching by year, tenders with visible amounts}\label{tab:chile_anio}
\input{tablas/chile_anio.tex}
\end{table}

\begin{figure}[H]\centering
\includegraphics[width=\textwidth]{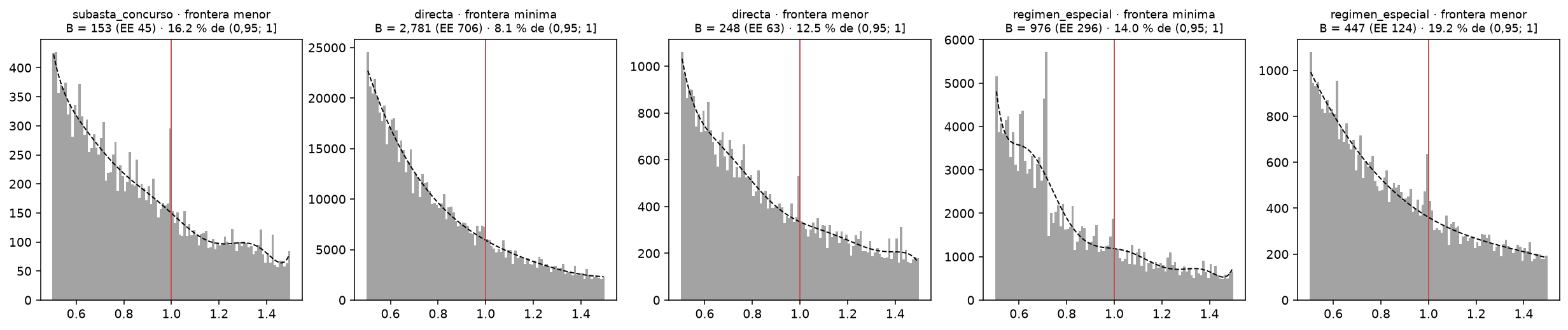}
\caption{Placebo densities. Samples without a value cap at the cutoff examined; same estimator as Figure~\ref{fig:principal}.}\label{fig:placebos}
\end{figure}

\begin{figure}[H]\centering
\includegraphics[width=0.8\textwidth]{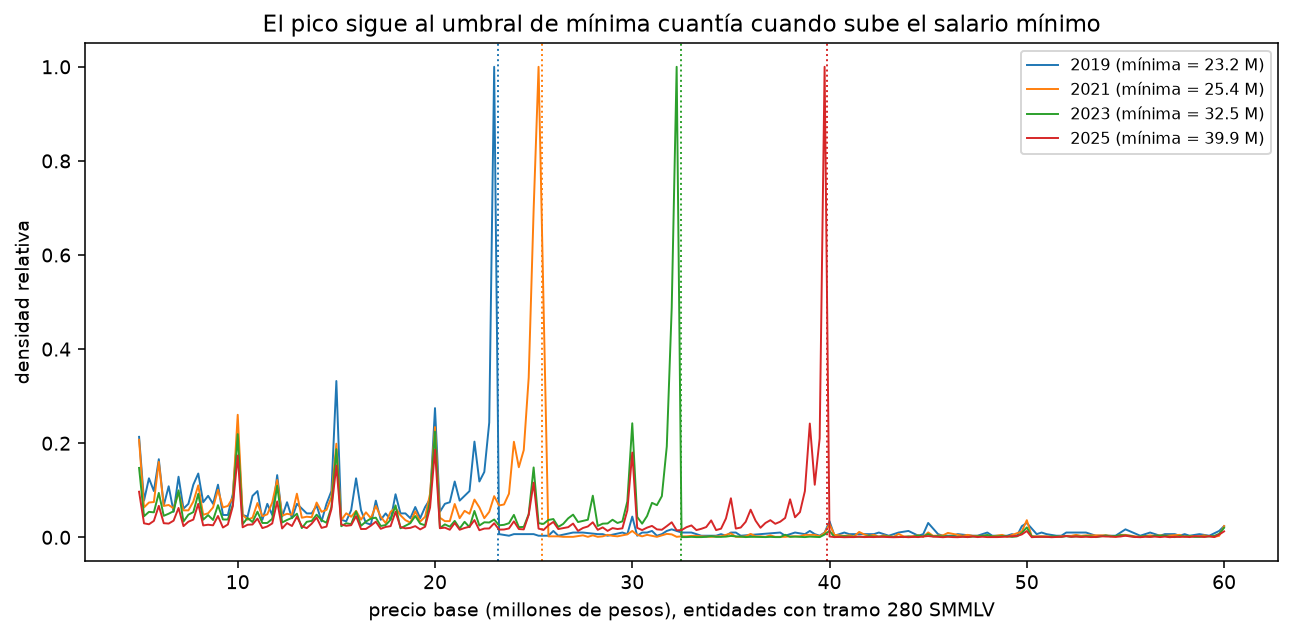}
\caption{The spike follows the minimum wage. Density of estimated values in pesos for entities in the 280 SMMLV band, selected years; dotted lines mark each year's small-purchase threshold.}\label{fig:movimiento}
\end{figure}

\begin{figure}[H]\centering
\includegraphics[width=0.7\textwidth]{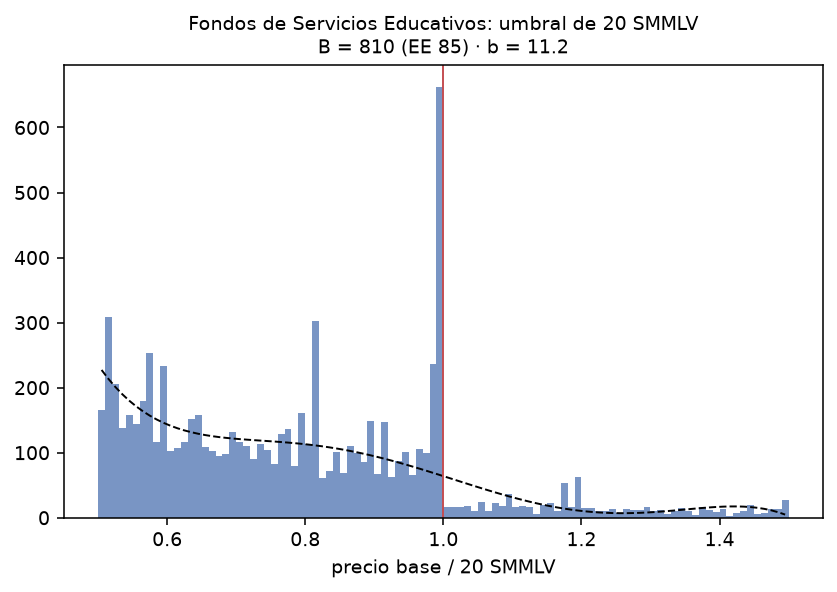}
\caption{Public schools: bunching below the 20 SMMLV limit of the school-fund regime.}\label{fig:fse}
\end{figure}

\section{Electoral restrictions: further episodes}\label{app:lg}

Article~38 of Law 996 bars territorial governments from signing intergovernmental agreements in the four months before local elections. Applying the event-study design to agreements signed by territorial entities around the 2019 and 2023 local elections (SECOP~II) gives a change of \LGartThreeEightConveniosTerritorialesVeda{} during the restriction ($p = \LGartThreeEightConveniosTerritorialesVedaP$), against \LGartThreeEightConveniosNacionalesVeda{} for national entities ($p = \LGartThreeEightConveniosNacionalesVedaP$). Between 12 November 2021, when Law 2159 lifted the restriction for the following year, and 28 January 2022, when the presidential ban took over, agreements between the Nation and territorial entities could be signed freely; a difference-in-differences against territorial-to-territorial agreements and against the same weeks of the 2017--18 cycle gives \PtresDD{} log points ($p = \PtresP$ from placebo windows in other years; the 2017--18 cycle alone gives \PtresDDctrl). Neither episode shows a detectable effect.

\begin{figure}[H]\centering
\includegraphics[width=\textwidth]{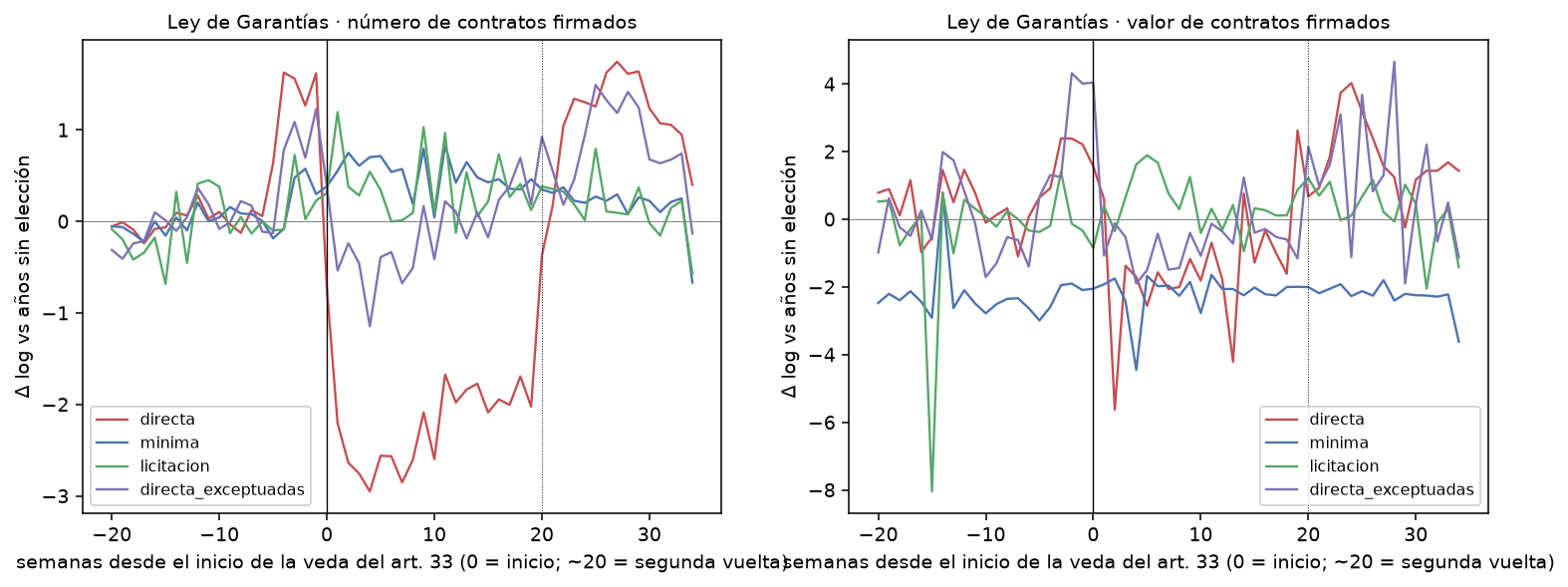}
\caption{SECOP~II only, cycles 2018, 2022 and 2026: difference in log weekly contracts between presidential and non-election years, by procedure and for the exempted sectors.}\label{fig:garantias_ii}
\end{figure}

%% file: tablas/robustez.tex
\begin{tabular}{lrrrrr}
\toprule
Specification & $z_U$ = 1.02 & $z_U$ = 1.04 & $z_U$ = 1.10 & $z_U$ by integration &  \\
\midrule
\textit{Small-purchase threshold (10\% of MC)} &  &  &  &  &  \\
degree 3, $z_L$ = 0.90 & 58,329 & 57,934 & 56,930 & 55,071 &  \\
degree 3, $z_L$ = 0.95 & 51,730 & 51,516 & 50,946 & 49,688 &  \\
degree 5, $z_L$ = 0.90 & 57,455 & 56,800 & 55,214 & 52,379 &  \\
degree 5, $z_L$ = 0.95 & 51,143 & 50,784 & 49,853 & 47,690 &  \\
degree 7, $z_L$ = 0.90 & 57,216 & 56,339 & 54,308 & 49,973 &  \\
degree 7, $z_L$ = 0.95 & 50,814 & 50,318 & 49,069 & 45,854 &  \\
degree 8, $z_L$ = 0.90 & 57,921 & 56,810 & 53,253 & 47,593 &  \\
degree 8, $z_L$ = 0.95 & 50,797 & 50,063 & 47,944 & 45,446 &  \\
degree 9, $z_L$ = 0.90 & 56,176 & 54,971 & 52,359 & 50,365 &  \\
degree 9, $z_L$ = 0.95 & 50,275 & 49,631 & 48,120 & 45,463 &  \\
degree 10, $z_L$ = 0.90 & 57,656 & 56,679 & 54,291 & 51,215 &  \\
degree 10, $z_L$ = 0.95 & 50,760 & 49,978 & 47,935 & 45,295 &  \\
\textit{Tender threshold (MC)} &  &  &  &  &  \\
degree 3, $z_L$ = 0.90 & 3,838 & 3,788 & 3,661 & 3,401 &  \\
degree 3, $z_L$ = 0.95 & 3,255 & 3,229 & 3,162 & 3,010 &  \\
degree 5, $z_L$ = 0.90 & 3,729 & 3,651 & 3,462 & 3,137 &  \\
degree 5, $z_L$ = 0.95 & 3,206 & 3,167 & 3,067 & 2,867 &  \\
degree 7, $z_L$ = 0.90 & 3,801 & 3,709 & 3,497 & 2,951 &  \\
degree 7, $z_L$ = 0.95 & 3,220 & 3,170 & 3,049 & 2,783 &  \\
degree 8, $z_L$ = 0.90 & 3,800 & 3,658 & 3,172 & 2,545 &  \\
degree 8, $z_L$ = 0.95 & 3,190 & 3,110 & 2,883 & 2,664 &  \\
degree 9, $z_L$ = 0.90 & 3,580 & 3,427 & 3,061 & 2,683 &  \\
degree 9, $z_L$ = 0.95 & 3,133 & 3,063 & 2,902 & 2,771 &  \\
degree 10, $z_L$ = 0.90 & 3,810 & 3,707 & 3,441 & 2,878 &  \\
degree 10, $z_L$ = 0.95 & 3,229 & 3,157 & 2,976 & 2,760 &  \\
\bottomrule
\end{tabular}

%% file: tablas/secop1.tex
\begin{tabular}{llrrrrr}
\toprule
Period & Threshold & $B$ & (s.e.) & Share & $b$ & $N$ \\
\midrule
2011-2020 & Small purchase & 149,534 & (6,702) & 83\% & 25.2 & 488,908 \\
2011 & Small purchase & 9,861 & (624) & 86\% & 30.4 & 39,967 \\
2012 & Small purchase & 11,812 & (692) & 72\% & 12.7 & 55,064 \\
2013 & Small purchase & 15,317 & (670) & 81\% & 21.8 & 56,121 \\
2014 & Small purchase & 20,179 & (1,012) & 91\% & 49.7 & 59,197 \\
2015 & Small purchase & 18,959 & (869) & 87\% & 33.3 & 57,457 \\
2016 & Small purchase & 17,350 & (697) & 92\% & 58.1 & 48,067 \\
2017 & Small purchase & 19,219 & (754) & 92\% & 55.1 & 49,789 \\
2018 & Small purchase & 13,350 & (534) & 78\% & 17.4 & 47,889 \\
2019 & Small purchase & 15,639 & (593) & 90\% & 43.1 & 46,183 \\
2020 & Small purchase & 12,955 & (522) & 93\% & 66.8 & 29,174 \\
2011-2020 & Tender & 4,956 & (301) & 56\% & 6.3 & 88,927 \\
2011 & Tender & 694 & (96) & 65\% & 9.5 & 9,434 \\
2012 & Tender & 377 & (47) & 57\% & 6.6 & 7,220 \\
2013 & Tender & 453 & (48) & 51\% & 5.3 & 10,023 \\
2014 & Tender & 551 & (61) & 54\% & 5.9 & 10,903 \\
2015 & Tender & 723 & (55) & 59\% & 7.2 & 12,182 \\
2016 & Tender & 431 & (41) & 61\% & 7.7 & 7,381 \\
2017 & Tender & 621 & (47) & 62\% & 8.2 & 9,259 \\
2018 & Tender & 353 & (36) & 48\% & 4.6 & 8,780 \\
2019 & Tender & 612 & (47) & 60\% & 7.5 & 9,700 \\
2020 & Tender & 416 & (28) & 74\% & 13.9 & 4,045 \\
\bottomrule
\end{tabular}

%% file: tablas/chile_anio.tex
\begin{tabular}{llrrrrr}
\toprule
Year & Threshold & $B$ & (s.e.) & Share & $b$ & $N$ \\
\midrule
2015 & 100 UTM & 1,576 & (348) & 59\% & 7.3 & 18,105 \\
2016 & 100 UTM & 1,693 & (256) & 70\% & 11.9 & 15,827 \\
2017 & 100 UTM & 1,709 & (363) & 74\% & 14.0 & 14,723 \\
2018 & 100 UTM & 1,498 & (333) & 70\% & 11.6 & 15,032 \\
2019 & 100 UTM & 2,813 & (803) & 83\% & 24.8 & 17,587 \\
2020 & 100 UTM & 2,231 & (370) & 79\% & 19.2 & 13,472 \\
2021 & 100 UTM & 2,768 & (383) & 74\% & 14.0 & 21,679 \\
2022 & 100 UTM & 2,946 & (482) & 68\% & 10.5 & 28,556 \\
2023 & 100 UTM & 4,561 & (651) & 76\% & 15.7 & 32,205 \\
2024 & 100 UTM & 4,679 & (698) & 75\% & 14.9 & 34,441 \\
2025 & 100 UTM & 1,188 & (265) & 74\% & 14.1 & 10,397 \\
2026 & 100 UTM & 739 & (195) & 71\% & 12.2 & 6,900 \\
2015 & 1,000 UTM & 452 & (145) & 56\% & 6.3 & 7,215 \\
2016 & 1,000 UTM & 461 & (117) & 64\% & 8.8 & 5,756 \\
2017 & 1,000 UTM & 495 & (91) & 66\% & 9.8 & 5,820 \\
2018 & 1,000 UTM & 490 & (104) & 66\% & 9.9 & 5,773 \\
2019 & 1,000 UTM & 385 & (91) & 59\% & 7.3 & 5,980 \\
2020 & 1,000 UTM & 531 & (99) & 65\% & 9.3 & 6,119 \\
2021 & 1,000 UTM & 547 & (102) & 57\% & 6.7 & 7,825 \\
2022 & 1,000 UTM & 367 & (63) & 47\% & 4.4 & 8,807 \\
2023 & 1,000 UTM & 593 & (89) & 57\% & 6.7 & 9,287 \\
2024 & 1,000 UTM & 567 & (104) & 58\% & 6.8 & 9,140 \\
2025 & 1,000 UTM & 558 & (100) & 58\% & 6.9 & 9,827 \\
2026 & 1,000 UTM & 542 & (126) & 66\% & 9.6 & 6,726 \\
\bottomrule
\end{tabular}